\documentclass[12pt,preprint]{aastex62}
\usepackage[english]{babel}
\usepackage{booktabs}
\usepackage{amsmath,amssymb}
\usepackage{amsthm}
\usepackage{graphicx}
\usepackage{braket}
\usepackage[table]{xcolor}
\usepackage{etoolbox}
\usepackage{multirow}
\usepackage{hyperref} 
\usepackage{pgf} 
\definecolor{high}{HTML}{008f00}  
\definecolor{low}{HTML}{c42c00}  
\definecolor{mid}{HTML}{ffffff}

\newcommand{\tc}{\tau_c}
\newcommand{\feh}{\left[\mbox{Fe/H}\right]}
\newcommand{\teff}{T_\mathrm{eff}}
\newcommand{\bprp}{\left(G_\mathrm{BP} - G_\mathrm{RP}\right)_0}
\newcommand{\vconv}{v_\mathrm{conv}}
\newcommand{\sone}{\mathcal{S}_1}
\newcommand{\stwo}{\mathcal{S}_2}
\newcommand{\sthree}{\mathcal{S}_3}

\begin{document}

\title{Asteroseismic calibration of the Rossby number and its connection to the stellar dynamo and fundamental properties}
\author{Alfio Bonanno}
\affiliation{INAF-Osservatorio Astrofisico di Catania, via S. Sofia 81, Catania, Italy}
\affiliation{Leibniz-Institut f\"{u}r Astrophysik Potsdam (AIP), An der Sternwarte 16, 14482 Potsdam, Germany}
\author{Enrico Corsaro}
\affiliation{INAF-Osservatorio Astrofisico di Catania, via S. Sofia 81, Catania, Italy}
\author{Travis S. Metcalfe}
\affiliation{Center for Solar-Stellar Connections, WDRC, 9020 Brumm Trail, Golden, CO 80403, USA}
\author{Sylvain Breton}
\affiliation{INAF-Osservatorio Astrofisico di Catania, via S. Sofia 81, Catania, Italy}
\author{Orlagh L. Creevey}
\affiliation{Université C\^{o}te d’Azur, Observatoire de la C\^{o}te d’Azur, CNRS, Laboratoire Lagrange, France}
\author{Christopher J. Lindsay}
\affiliation{Department of Astronomy, Yale University, P.O. Box 208101, New Haven, CT 06520-8101, USA}
\begin{abstract}
   
The stellar Rossby number, a dimensionless parameter quantifying the influence of Coriolis forces on convective motions, plays a pivotal role in understanding magnetic stellar evolution. In this work, we explore the connection between the Rossby number and potential dynamo mechanisms in Sun-like stars, as well as its dependence on fundamental stellar properties. We present a novel, detailed asteroseismic calibration of the convective turnover time, incorporating for the first time Gaia photometry alongside surface gravity, effective temperature, and stellar metallicity. Our analysis employs an expanded sample of more than 150 stars, including targets from the Kepler LEGACY and KOI surveys, as well as more evolved stars observed by TESS and K2. This sample spans evolutionary stages from the main sequence to the early red giant branch (RGB), enabling a comprehensive investigation of Rossby number trends across stellar evolution.

\end{abstract}

\section{Introduction}
A crucial quantity for understanding how rotation modifies fluid turbulence is the  Rossby number, defined as $Ro = u' / \Omega \ell$, where $u'$ represents turbulent velocity fluctuations and $\ell$ is the characteristic eddy size. The physical meaning of this parameter is straightforward: when $Ro \ll 1$, the Coriolis force dominates over turbulent forces, whereas for $Ro \gg 1$, turbulence is largely unaffected by large-scale rotation.

In the context of low-mass stars, $\Omega$ denotes the stellar rotation rate, and $u' / \ell \approx \tc^{-1}$ (where $\tc$ is the convective eddy turnover time). In particular the large scale magnetic field observed in low-mass stars is believed to arise from a dynamo mechanism operating in the convection zone, and for this reason various studies have attempted to correlate $Ro$ with magnetic activity indicators.  A well-established correlation, known since 
 the work of \citet{1981ApJ...248..279P},  connects  rotation with X-ray luminosity and presents two different scaling laws at a critical Rossby number \citep{2003A&A...397..147P,2011ApJ...743...48W,2022A&A...662A..41R} whose physical meaning is still much debated.  In contrast, the relationship between the Rossby number and a chromospheric activity \cite{1984ApJ...279..763N}
 diagnostic like the Mt. Wilson $S$-index or  the fractional  Ca II HK emission flux above the backround $\langle R'_{\text{HK}} \rangle$ turned out to be more elusive, with scaling laws which are possibly different from the one determined from the X-ray luminosity \citep{2008ApJ...687.1264M,2020NatAs...4..658L}. One additional difficulty  in relating the rotation rate with the activity cycle lies in the presence of distinct ``branches" in the $\log P_{\text{rot}} / P_{\text{cyc}}$ vs. $\log \langle R'_{\text{HK}} \rangle$ diagram \citep{2007ApJ...657..486B,1999ApJ...524..295S,Bonanno14}, which may reflect different dynamo mechanisms operating in F, G, K, and M stars—thereby complicating the identification of a universal scaling law.  In recent times the discovery of weakened magnetic braking in old solar-type stars \citep{2016Natur.529..181V} has suggested the possibility of the existence of a critical Rossby number beyond which the dynamo's efficiency drops dramatically \citep[e.g., see][]{Metcalfe2025}. These findings underscore the need to clarify the link between the Rossby number and the underlying dynamo processes in stellar interiors.  

In solar-like stars, a major challenge in defining a robust prescription for the Rossby number arises from the steep variation of the convective turnover time ($\tc$) with depth near the base of the convection zone. Although low-mass main-sequence stars such as the Sun exhibit differential rotation both internally and at the surface, the degree of non-uniformity is relatively moderate. For example, in the Sun, the latitudinal differential rotation across the whole convection zone is present but not particularly strong. However, standard mixing-length theory predicts a sharp {increase} in $\tc$ to timescales of months near the base of the solar convection zone—precisely the region where the giant convective cells are believed to form \citep{2013Sci...342.1217H}. As a result, defining the Rossby number becomes problematic, since the quantity formally diverges at the base of the convection zone, where the stellar dynamo is thought to operate (\citealt{2002A&A...390..673B}, but see also \citealt{Gossage24} for more discussion).

To address this issue, a recent study \citep{Corsaro21} proposed defining the Rossby number as $Ro = P_{\rm rot} / \tc$, where the convective turnover time is estimated as
\begin{equation}
\label{t1}
    \tc = \frac{d_{\rm CZ}}{\langle v \rangle},
\end{equation}
with $d_{\rm CZ}$ denoting the thickness of the convection zone and $\langle v \rangle \approx (L R / M)^{1/3}$ representing the average convective velocity \citep[see also][]{2017LRSP...14....4B}. Here, $L$, $R$, and $M$ correspond to the stellar luminosity, radius, and mass, respectively. 

It is reassuring to note that Eq.~(\ref{t1}) provides a good approximation of the local convective turnover time at the base of the convection zone as inferred from stellar models and MLT theory. For instance, in the case of the Sun, one obtains $\tc \approx 45$ days, a value consistent with the local convective turnover time at $0.72\,R_\odot$ in a fully calibrated solar standard model \citep{2002A&A...390.1115B}, in agreement with the analysis of \cite{2010A&A...510A..46L}.
The key parameter in Eq.~(\ref{t1}) is $d_{\rm CZ}$, a quantity that can be reliably estimated via asteroseismology for a carefully selected sample of well-characterized stars, for which internal structural profiles can be determined with high confidence, as in \cite{Corsaro21}.

In the present study, we extend that analysis by incorporating Gaia photometry along with surface gravity, effective temperature, and stellar metallicity. Our updated approach uses an expanded sample of more than 150 stars, including targets from the \textit{Kepler} LEGACY and KOI surveys, as well as more evolved stars observed by \textit{TESS} and \textit{K2}. This dataset spans evolutionary stages from the main sequence to the early red giant branch (RGB), {covering almost 2000\,K in temperature, $4800\,K \lesssim \teff \lesssim 6700\,K$, and a range in Gaia color index of $0.55 \lesssim \bprp \lesssim 1.25$,} allowing for a comprehensive investigation of Rossby number trends across stellar evolution.

We describe the sample selection and its observational and modelled properties in Section~\ref{sec:sample}. Section~\ref{sec:multilinear} introduces the multi-linear relationships that we investigate through the adoption of a Bayesian framework. The results of the analysis are shown and detailed in Section~\ref{sec:results}. Finally, we provide a comprehensive discussion of our findings in Section~\ref{sec:discussion} in relation to the effect of the dynamo action, and we draw our conclusions in Section~\ref{sec:conclusions}.

\section{The sample}
\label{sec:sample}
In this work, we considered a cohort of 97 stars from the LEGACY \citep{Lund_2017} and KOI \citep{2016MNRAS.456.2183D} samples observed with \textit{Kepler}, consisting of main-sequence and early subgiant stars (hereafter sample $\mathcal{S}_1$). Additionally, we analyzed a second cohort of 62 stars \citep{2024ApJ...965..171L}, including 36 observed with \textit{Kepler}, eight with K2, and 18 with TESS, representing late subgiants and early red giant branch (RGB) stars (hereafter sample $\mathcal{S}_2$). We also included the Sun as an additional reference point in sample $\mathcal{S}_1$ for subsequent analysis.  Furthermore, we considered a third case, a combined sample $\mathcal{S}_3 \equiv \mathcal{S}_1 \cup \mathcal{S}_2$, to explore the feasibility of calibrating a unified relationship for $\tc$ that spans multiple evolutionary stages.  Estimates of $T_\mathrm{eff}$, $\log g$, and $\left[ \mbox{Fe/H} \right]$ are generally derived from spectroscopic observations. Specifically, $\log g$ values for $\mathcal{S}_1$ were obtained by combining multiple spectroscopic sources available in the PASTEL catalog \citep{Soubiran10PASTEL} using a weighted averaging approach. The adopted color index, $\bprp$, with its corresponding reddening correction applied, was taken from Gaia DR3 for most stars in the sample. For stars not having a released Gaia DR3 reddening correction, we evaluated the reddening in  ($B-V$) color band through the adoption of the STILISM 3D dust maps \citep{Lallement18}, which we then converted into a reddening in the Gaia color band following \cite{Casagrande18bol_corr}. We note that, since the stars considered in this analysis are relatively bright, in most cases the reddening values appear to be rather small, on the order of $10^{-2}$--$10^{-3}$ mag.

\subsection{Stellar modeling of the targets}
All the stars in this work have been modeled by exploiting their detailed asteroseismic content. In particular, we used version 1.3 of the Asteroseismic Modeling Portal \citep[AMP;][]{AMP2009} for the stars in $\sone$ \citep[for details, see][]{Creevey2017}, to obtain $M$, $R$, the thickness of the convection zone $d_\mathrm{CZ}$, and $L$, which are then used for the evaluation of the convective turnover time $\tc$ and of the convective velocity $v_\mathrm{conv}$ following the derivation presented in \cite{Corsaro21} (their Eq.s~(5) and (6), respectively, see also Sect.~\ref{sec:multilinear}). The stars in $\stwo$ were instead modeled by \cite{2024ApJ...965..171L}, with $M$, $R$, $L$ already published, while $d_\mathrm{CZ}$ is provided in this work, and $\tc$ and $v_\mathrm{conv}$ are once again evaluated following \cite{Corsaro21}.

\section{Multi-linear relationships for $\tc$}
\label{sec:multilinear}
In the work done by \cite{Corsaro21} the authors analyzed simple linear and quadratic relations connecting $\tc$ with the color index of the star. A limitation of the relations considered is that they do not take into account other stellar atmospheric properties such as surface gravity and metallicity, which could play an important role in influencing the efficiency of convection (e.g. see \citealt{Bonanno22}). In order to preserve a proper analytical treatment of the error propagation, one can rely on multi-linear relations that link multiple observables to the same quantity for which a prediction is needed, similarly to the work done by \cite{Corsaro13}. 
Therefore, in this section we will analyze relationships connecting $\tc$ to multiple observables related to measurable stellar atmospheric properties, thus expanding upon the analysis done previously by \cite{Corsaro21}. This can allow us not only to improve the quality (in particular the accuracy) of the  predictions but also to gain a better understanding and a quantification of the impact these atmospheric properties have on a characteristic time scale of stellar convective motion. In particular, we start by considering a multi-power relation of the type
\begin{equation}
\tc \left( \alpha_1, \alpha_2, \dots, \alpha_N, \beta \right)  = \beta \prod_{i=1}^{N} O_i^{\alpha_i} \, ,
\end{equation}
where $O_i$, for $i=1, \dots, N$, are the observables, while $\alpha_i$ are the corresponding exponents that, along with the multiplication factor $\beta$, have to be estimated from a fit. Following the approach presented by \cite{Corsaro13} (but see also \citealt{Bonanno14} and \citealt{Corsaro_2017} for additional applications), we can linearize the power-law relation to become
\begin{equation}
\ln \tc \left( \alpha_1, \alpha_2, \dots, \alpha_N, \beta \right) = \ln \beta + \sum_{i=1}^N 	\alpha_i \ln O_i \, .
\label{eq:multi-linear}
\end{equation}
In our application, the observables $O_i$ that are taken into account are: $\bprp$, $\log g$, $\feh$, $\teff$, thus a total of N = 4 components of the multi-linear relation. Despite $\bprp$ also {containing} information about the temperature of the star, it is a physically different quantity than $\teff$, suggesting that it is still worth to investigate whether $\teff$ alone is playing any significant role on top of the color index in predicting the value of $\tc$.

\begin{figure}[ht]
   \centering
  \includegraphics[width=18cm]{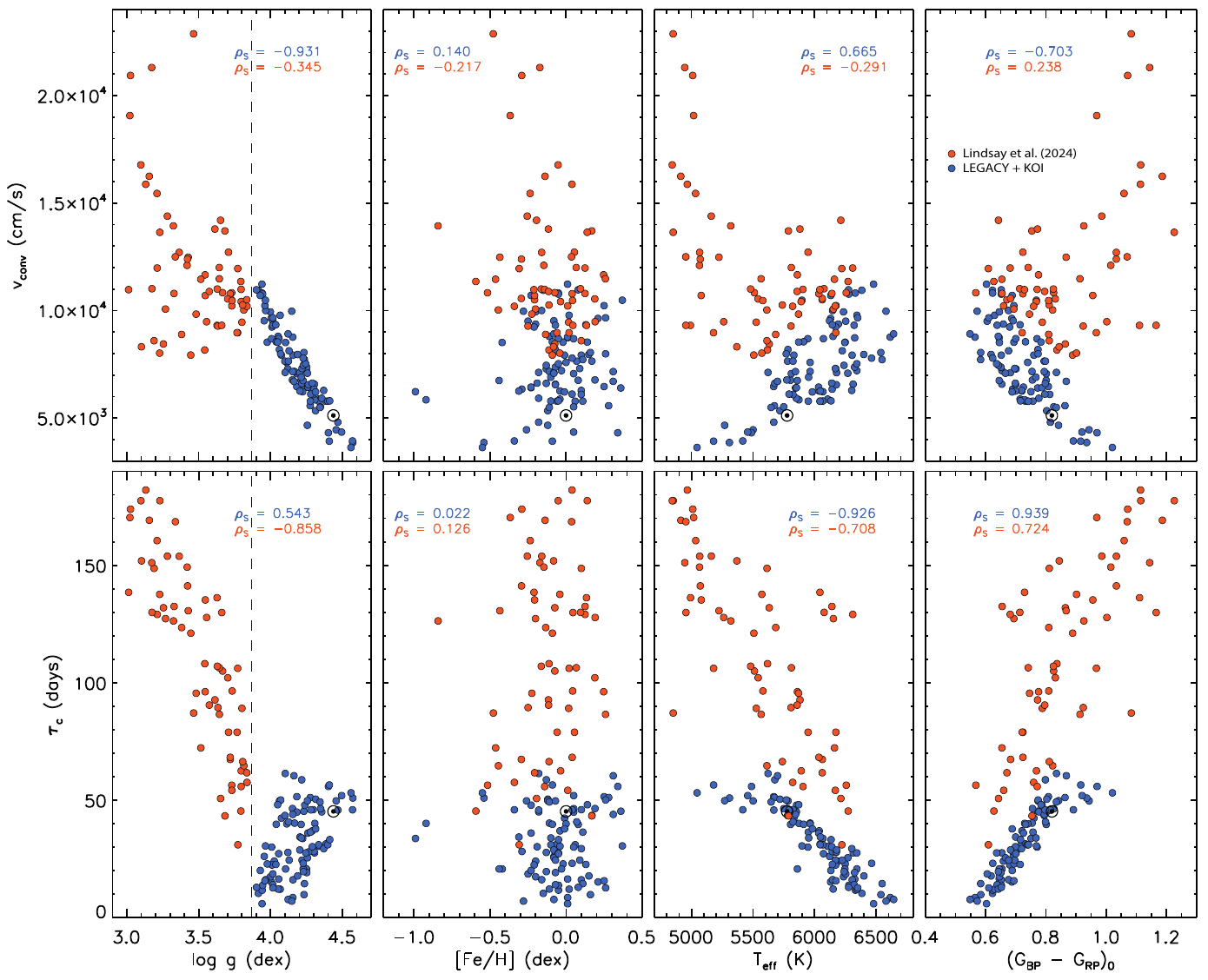}
    \caption{Convective velocities $\vconv$ and convective turnover time $\tc$ as a function of Gaia DR3 color index $\bprp$, surface gravity $\log g$, metallicity $\feh$, and temperature $\teff$ for both samples (main sequence and early subgiants from LEGACY + KOI in blue, late subgiants and early RGB from \cite{2024ApJ...965..171L} orange). Spearman's rank correlation coefficients are also indicated for each sample using the same color coding. The dashed vertical line represents the cut in $\log g$ applied to distinguish $\sone$ from $\stwo$, as described in Sect.~\ref{sec:discussion}.}
    \label{fig:correlations}
\end{figure}

However, before performing an actual fit to this relation, it is useful to inspect the potential correlations that could arise between $\tc$ and each of these observables. In Fig.~\ref{fig:correlations} we show the result of these correlations for both $\tc$ and $\vconv$, the latter being defined as $(LR/M)^{1/3}$, for each of the two samples considered. One can see clear correlations with most of the observables but a less clear one seems to be present for the case of metallicity $\feh$, whose statistical significance ought to be investigated. The evaluation of Spearman's rank coefficients already provides a hint of the strength and kind of correlations that can be found. We also notice that for more evolved stars (orange bullets) the correlations appear more loose, as a direct consequence of the higher sensitivity of stellar evolution to stellar mass and luminosity.

The fits are performed by {means of the public Bayesian inference tool \textsc{DIAMONDS}\footnote{The software can be downloaded from its official GitHub page available at https://github.com/EnricoCorsaro/DIAMONDS .}, with uniform (flat) priors on each model free parameter. Uniform prior ranges are adjusted to ensure that the global maximum of the likelihood distribution is fully resolved within the inspected parameter space. We adopt} a specific Likelihood function that is formally equivalent to the standard Normal Likelihood, except for the treatment of the uncertainties. As explained in \cite{Corsaro13}, the uncertainty on each data point is here evaluated as a total uncertainty arising from that of each of the observables (or independent variables) taken into account, including the one on the dependent variable $\tc$. The uncertainty on a single data point $j$ thus depends on the model free parameters $\{ \alpha_i \}$ and it is evaluated as
\begin{equation}
\widetilde{\sigma_j}^2 \left( \alpha_1, \alpha_2, \dots, \alpha_N \right) = \widetilde{\sigma}_{\tc,j}^2 + \sum_{i=1}^{N} \alpha_i^2 \widetilde{\sigma}_{\mathrm{O},i,j}^2
\end{equation}
where  $\widetilde{\sigma}$ is the relative uncertainty as we are dealing with logarithmic quantities in the linearized model (e.g. $\widetilde{\sigma}_\mathrm{\tc} \equiv \sigma_\mathrm{\tc} / \tc$). Our likelihood is therefore a multi-variate likelihood that also depends on the models' free parameters. 

\subsection{$\tc$ vs $\bprp$, $\log g$, $\feh$, $\teff$}
The general multi-linear model considered in our investigation is expressed in the linearized form as
\begin{equation}
\ln \tc \left( \alpha_1, \alpha_2, \alpha_3, \alpha_4, \beta \right) = \ln 	\beta + \alpha_1 \bprp + \alpha_2 \log g + \alpha_3 \feh + \alpha_4 \ln T_\mathrm{eff} \,.
\label{eq:multi-linear_2}
\end{equation}
where the terms for color index $(G_\mathrm{BP} - G_\mathrm{RP})_0$, surface gravity $\log g$, and metallicity $\mbox{[Fe/H]}$, are appearing in linear scale because they already represent logarithmic quantities. The total relative uncertainty is given by
\begin{equation}
\widetilde{\sigma_j}^2 \left( \alpha_1, \alpha_2, \alpha_3, \alpha_4 \right) = \widetilde{\sigma}_{\tc,j}^2 + \alpha_2^2 \sigma_{\log g, j}^2 + \alpha_3^2 \sigma_{\mathrm{[Fe/H]}, j}^2 + \alpha_4^2 \widetilde{\sigma}_{T_\mathrm{eff},j}^2 \, ,
\end{equation}
where the uncertainty on $\log g$ and $\feh$ is indicated as a standard uncertainty (and not relative) because both $\log g$ and $\feh$ are already logarithmic quantities (as stated above). Note that the coefficient $\alpha_1$ corresponding to the color index does not contribute to the total uncertainties because the uncertainties in the color index are set to zero by default (their contribution is negligible compared to that of the other observables). 

Given the number of observables to test (excluding the color index, which we consider as the most striking contribution, widely tested in previous literature works), we have a total of eight possible models to investigate, which have all been tested on the available datasets. To further motivate our adoption of multi-linear laws from a statistical point of view and to put them into context with our previous work, on top of these models we also performed the analysis of samples $\sone$, $\stwo$, and $\sthree$ using the quadratic law presented by \cite{Corsaro21} (their Eq.~(11), hereafter model $\mathcal{M}_\mathrm{quadratic}$). The multi-linear models are detailed in the following section. 

\section{Results}
\label{sec:results}
For each multi-linear model investigated, we generated plots to show how the predictions compare to the observations. For providing additional evidence, a panel with the residuals in logarithmic scale is included for each fit. The residuals are in turn fitted through a polynomial to quantify the presence of possible trends therein. For each model we evaluate the systematic deviation $s_\mathrm{dev}$ (in \%) from the expected pure match line (happening at 0) and standard deviation $\sigma_\mathrm{res}$ (also in \%) of the residuals. These two quantities give an indication of how much systematic deviation and dispersion, respectively, are present in the residuals of the selected model. The values of $s_\mathrm{dev}$ and $\sigma_\mathrm{res}$ can be easily compared to those of other models because they are indicated in relative percentage. Clearly, the larger the numbers, the worse the result of the fit. In addition, two additional quantities that are depending on the specific pair of observables considered, i.e. $D_\mathrm{avg}$ and $I_\mathrm{abs}$, are computed. The former is computed as the average of the absolute value of the derivative of the polynomial fit, while the latter is the integral of the absolute value of the polynomial fit. Both quantities are evaluated on a normalized observable-wise scale, so that they can be compared from pair to pair and also with respect to different models. In principle $D_\mathrm{avg}$ can provide an indication of the strength of nonlinear trends (i.e. curved dependencies) in the residuals, while $I_\mathrm{abs}$ can be seen as a net effect of how much the residuals deviate from a pure match between model predictions and observations (taking into account both dispersion and systematics). Again, higher values of $D_\mathrm{avg}$ and $I_\mathrm{abs}$ are indicative of a worse result. 

The details of the results for each model considered, taking into account the three samples $\mathcal{S}_1$, $\mathcal{S}_2$, and $\mathcal{S}_3$ are outlined in Sects. from~\ref{sec:m1} to \ref{sec:m8}.

\subsection{Model comparison}
For the sake of completeness, given that the models investigated here are linearly related to one another, one should take into account a multiplicity adjustment for the evaluation of the \emph{odds} ratio $\mathcal{O}$. Therefore for a proper assessment of the model comparison process, according to Bayesian statistics the odds ratio between two competing models $\mathcal{M}_i$ and $\mathcal{M}_j$ is defined as
\begin{equation}
\mathcal{O}_{ij} = \frac{\mathcal{E}_i}{\mathcal{E}_j} \frac{\pi (\mathcal{M}_i)}{\pi (\mathcal{M}_j)} = \mathcal{B}_{ij} \frac{\pi (\mathcal{M}_i)}{\pi (\mathcal{M}_j)} \, ,
\end{equation}
namely the product of the ratio of the models' evidences by the ratio of the corresponding models' priors. The models' priors taking into account the multiplicity adjustments can be evaluated according to 
\cite{Scott_2010}  (see also \citealt{Corsaro_2017}). We have that
\begin{equation}
\pi (\mathcal{M}_i) = \frac{k_i! (m - k_i)!}{m! (m + 1)} \, ,
\end{equation}
where $m$ is the maximum number of free parameters available in the model comparison process (here $m = 4$), while $k_i$ is the actual number of free parameters of the model $\mathcal{M}_i$. For the cases considered here we only have small correction factors, which are summarized below:
\begin{itemize}
\item $\pi (\mathcal{M})_{k=4} / \pi (\mathcal{M})_{k=1} = 4$
\item $\pi (\mathcal{M})_{k=4} / \pi (\mathcal{M})_{k=2} = 6$
\item $\pi (\mathcal{M})_{k=4} / \pi (\mathcal{M})_{k=3} = 4$
\item $\pi (\mathcal{M})_{k=3} / \pi (\mathcal{M})_{k=2} = 3/2$
\item $\pi (\mathcal{M})_{k=3} / \pi (\mathcal{M})_{k=1} = 1$
\end{itemize}
According to our definition of the models, we see that $\mathcal{M}_1$ has $k = 1$, $\mathcal{M}_2$, $\mathcal{M}_3$, and $\mathcal{M}_4$ have $k=2$, $\mathcal{M}_5$, $\mathcal{M}_6$, $\mathcal{M}_7$ have $k = 3$, and $\mathcal{M}_8$ has $k = 4$. Therefore, the multiplicity adjustment applied to the Bayes factor does not have a significant impact on the outcome of the model comparison process when evaluating the odds ratio $\mathcal{O}_{ij}$. We can conclude that $\mathcal{B}_{ij}$ alone suffices to draw conclusions on the model selection process.

The Bayes factors $\mathcal{B}_{ij}$ for each pair of models $(\mathcal{M}_i, \mathcal{M}_j)$ are listed in Tables~\ref{tab:s1}, \ref{tab:s2}, and \ref{tab:s3} for the samples $\sone$, $\stwo$, and $\sthree$, respectively. {Gray shading is used to highlight the favored models in the comparison.} The estimated values of the free parameters for the best-performing models, as determined through the Bayesian model comparison process, are presented in Table~\ref{tab:coeff}. {The corresponding correlation coefficients of the best models' free parameters, evaluated through the posterior probability distribution obtained from our inference, are presented in Table~\ref{tab:corr}.}

\subsection{Model $\mathcal{M}_1$ ($\alpha_1 \neq 0$, $\alpha_2 = \alpha_3 = \alpha_4 = 0$)}
\label{sec:m1}
\begin{figure}[h]
   \centering
  \includegraphics[width=18cm]{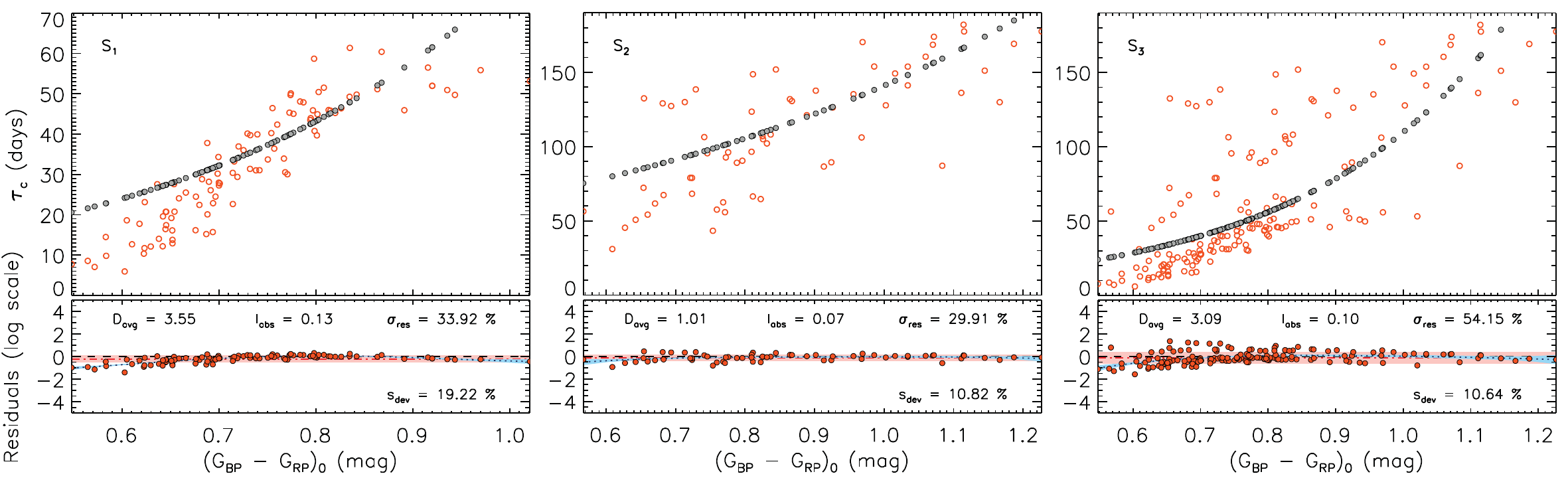}
    \caption{Dispersion plots showing the $\tc$ predictions from model $\mathcal{M}_1$. The data points are shown in color (open symbols), while the model predictions are represented by the gray (filled) bullets. A panel showing the residuals of the fit in the form $\ln \tc - \ln \tc^\mathrm{predict}$ is also shown for each observable. Also indicated are: a polynomial fit with its 1-$\sigma$ confidence region (dotted line with light blue shading), the average value of the residuals (dot-dashed red line), and the indices $\sigma_\mathrm{res}$, $s_\mathrm{dev}$, $D_\mathrm{avg}$, $I_\mathrm{abs}$.}
    \label{fig:m1}
\end{figure}

In this version of the model we consider only the dependency of $\tc$ with $\bprp$,
\begin{equation}
\ln \tc \left( \alpha_1, \beta \right) = \ln 	\beta + \alpha_1 \bprp \,.
\label{eq:m1}
\end{equation}
The results for this model are shown in Fig.~\ref{fig:m1} for all the samples. It is clear that the model is doing a rather poor job at matching the observations, in particular for reproducing the dispersion of the sample. The match between predictions and observations is better when considering the sample of evolved stars, and it becomes the worst when both $\sone$ and $\stwo$ are merged. The result gives already a strong indication that a simple linear relation on the sole color-index observable is not enough.

\subsection{Model $\mathcal{M}_2$ ($\alpha_1 \neq 0$, $\alpha_2 = \alpha_3 = 0$, $\alpha_4 \neq 0$)}
\begin{figure}[h]
   \centering
  \includegraphics[width=18cm]{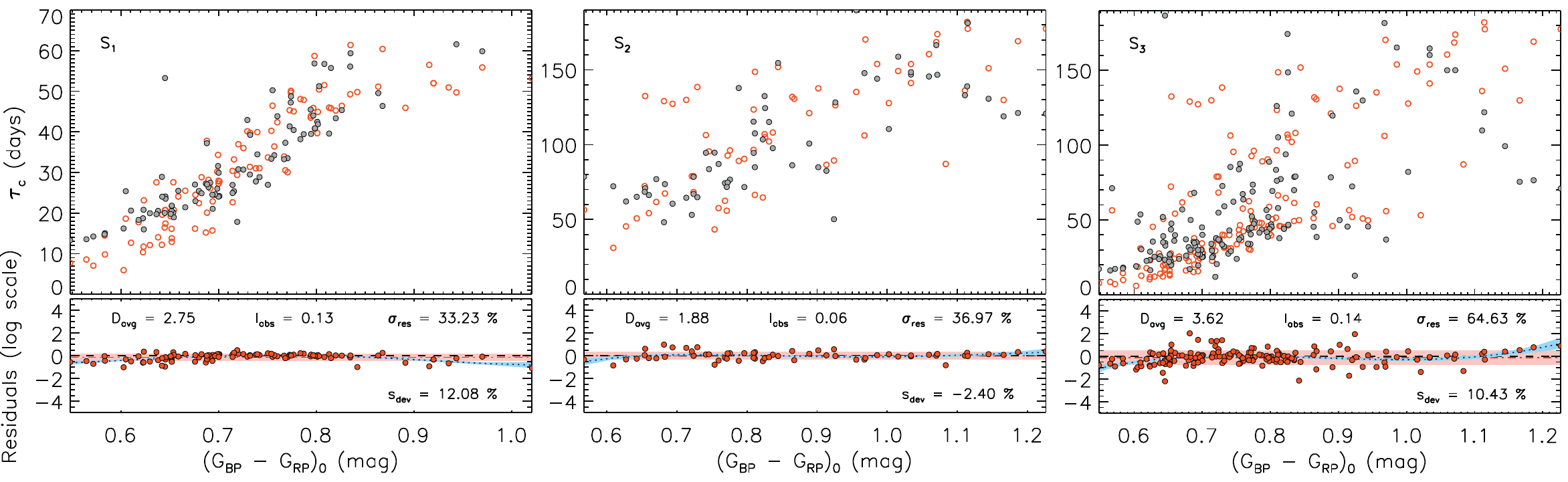}  
    \includegraphics[width=18cm]{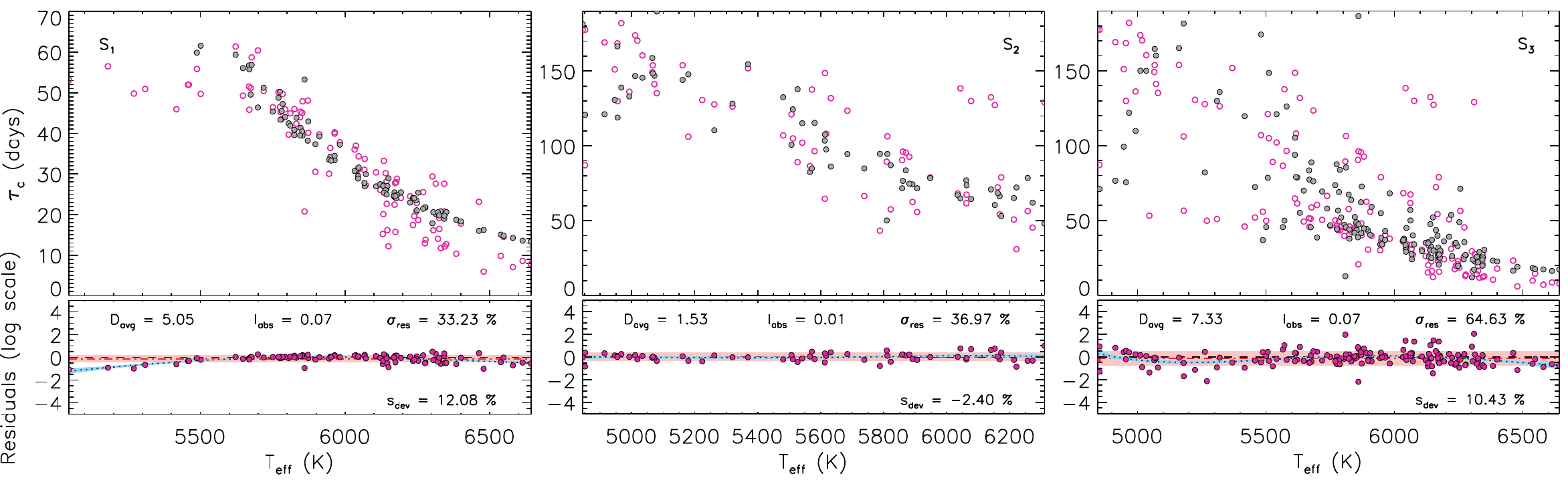}
    \caption{Similar description as for Fig.~\ref{fig:m1} but for model $\mathcal{M}_2$.}
    \label{fig:m2}
\end{figure}

\begin{equation}
\ln \tc \left( \alpha_1, \beta \right) = \ln 	\beta + \alpha_1 \bprp + \alpha_4 \ln T_\mathrm{eff}
\label{eq:m2}
\end{equation}
In this version of the model we consider the dependencies of $\tc$ with $\bprp$ and $\teff$. The results for the model are shown in Fig.~\ref{fig:m2} for all the samples. Despite a new observable is included with respect to model $\mathcal{M}_1$, there are visible systematic trends in the residuals of the fits to $\sone$ and $\sthree$. The situation is again significantly better when $\stwo$ alone is considered, while it is the worst with $\sthree$.

\subsection{Model $\mathcal{M}_3$ ($\alpha_1 \neq 0$, $\alpha_2 = \alpha_4 = 0$, $\alpha_3 \neq 0$)}
\begin{figure}[h]
   \centering
  \includegraphics[width=18cm]{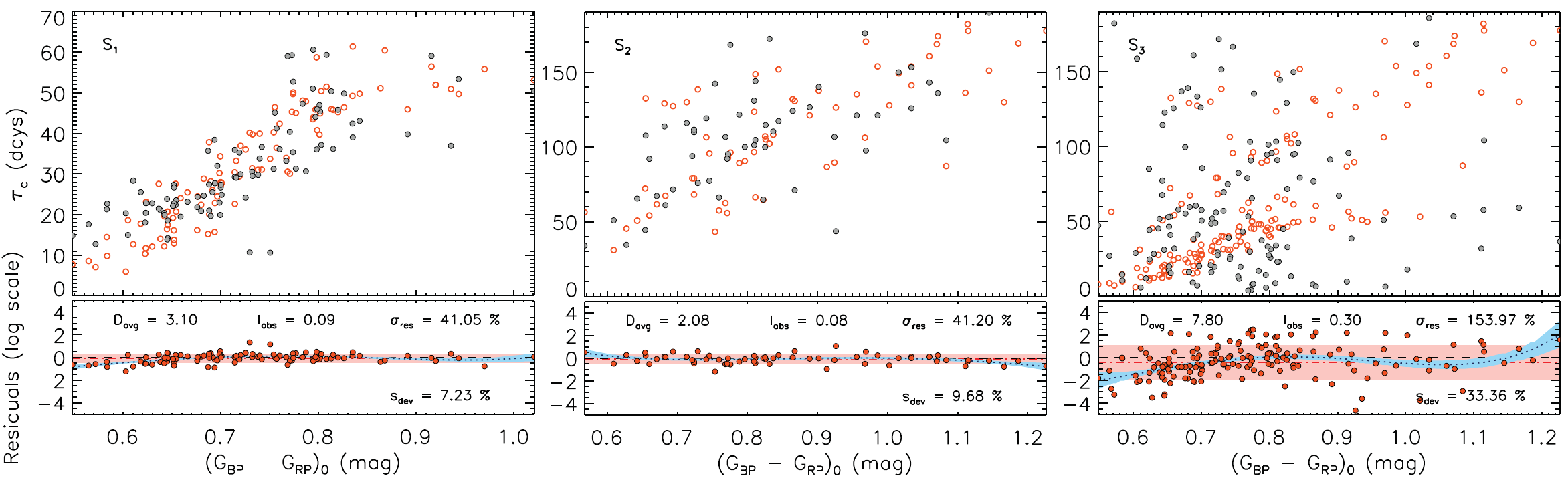}
  \includegraphics[width=18cm]{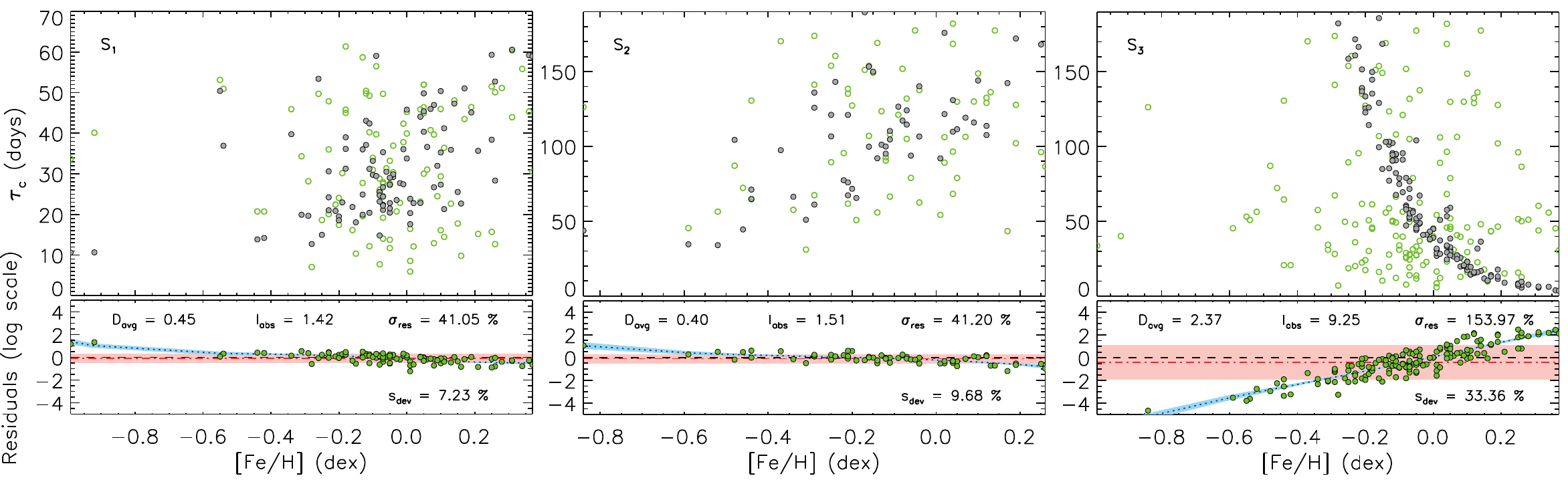}
    \caption{Similar description as for Fig.~\ref{fig:m1} but for model $\mathcal{M}_3$.}
    \label{fig:m3}
\end{figure}

\begin{equation}
\ln \tc \left( \alpha_1, \beta \right) = \ln 	\beta + \alpha_1 \bprp + \alpha_3 \feh
\label{eq:m3}
\end{equation}
In this version of the model we consider the dependencies of $\tc$ with $\bprp$ and $\feh$. The results for the model are shown in Fig.~\ref{fig:m3} for all the samples.  This model appears one of the worst at reproducing the results, especially for $\sthree$, where apparently the lack of a strong correlation between $\tc$ and $\feh$ is causing the overall fit to be less clear.

\subsection{Model $\mathcal{M}_4$ ($\alpha_1 \neq 0$, $\alpha_2 \neq 0$, $\alpha_3 = \alpha_4 = 0$)}
\begin{figure}[h]
   \centering
  \includegraphics[width=18cm]{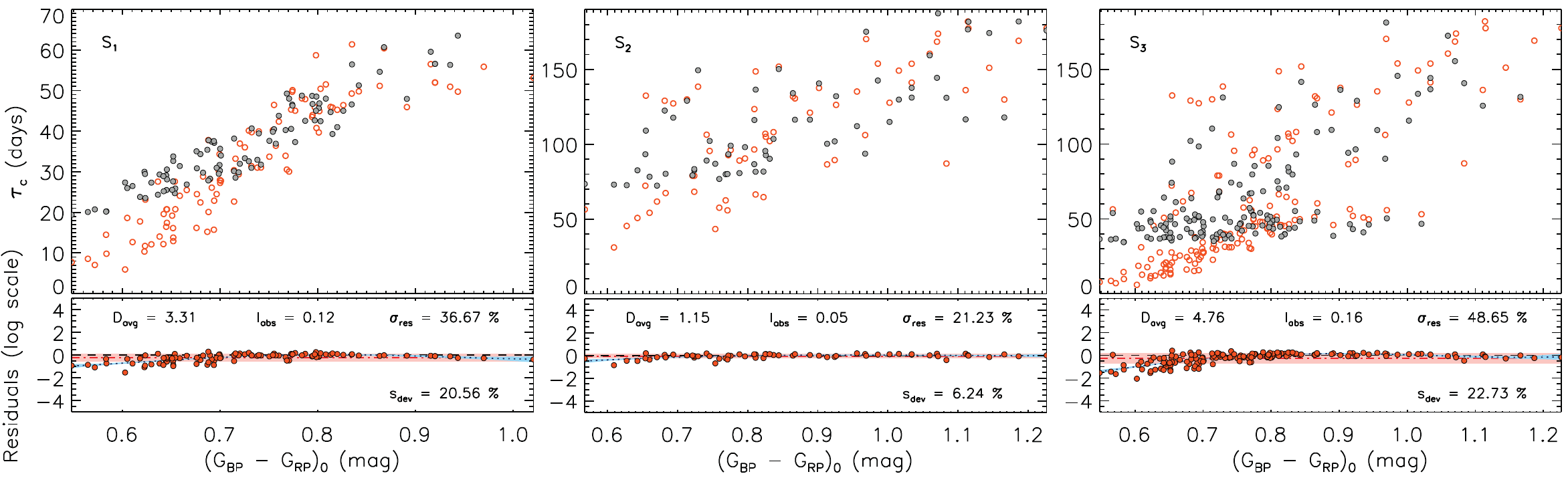}
  \includegraphics[width=18cm]{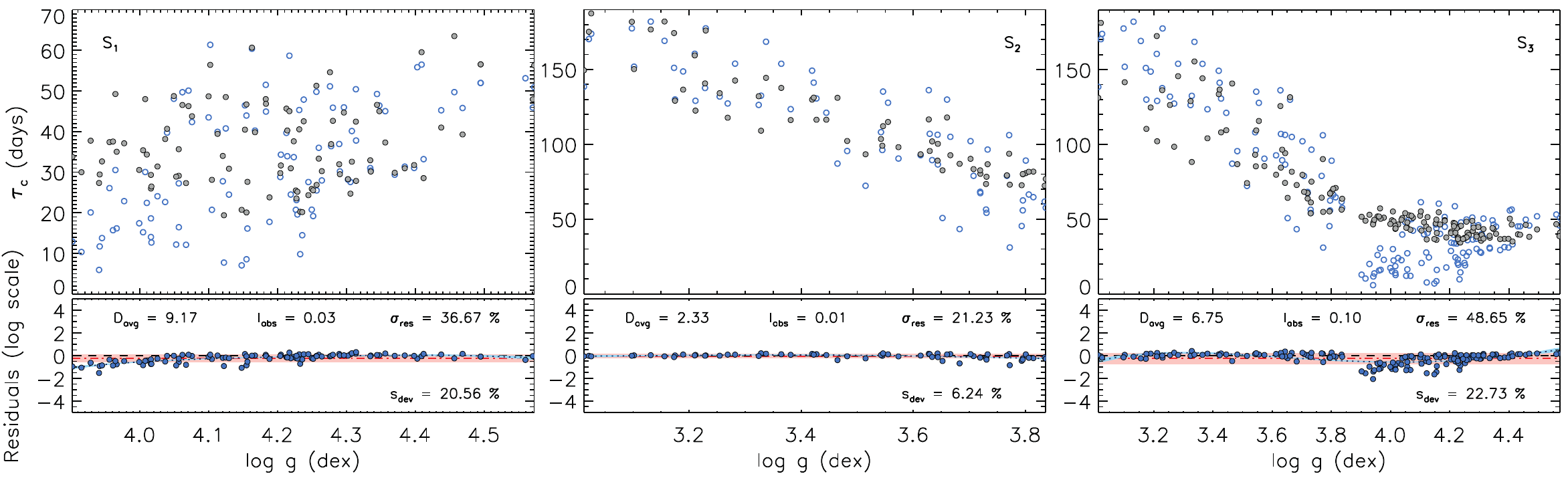}
    \caption{Similar description as for Fig.~\ref{fig:m1} but for model $\mathcal{M}_4$.}
    \label{fig:m4}
\end{figure}

\begin{equation}
\ln \tc \left( \alpha_1, \beta \right) = \ln 	\beta + \alpha_1 \bprp + \alpha_2 \log g
\label{eq:m4}
\end{equation}
In this version of the model we consider the dependency of $\tc$ with both $\bprp$ and $\log g$. The results for the model are shown in Fig.~\ref{fig:m4} for all the samples. While the general matching of the prediction is significantly improved with respect to model $\mathcal{M}_1$ at least in terms of observed dispersion of the sample, sensible trends are still visible in the residuals of the fit, as well as a similar systematic offset as the one found for model $\mathcal{M}_1$, now visible for both $\bprp$ and $\log g$. 

\subsection{Model $\mathcal{M}_5$ ($\alpha_1 \neq 0$, $\alpha_2 = 0$, $\alpha_3 \neq 0$, $\alpha_4 \neq 0$)}
\begin{equation}
\ln \tc \left( \alpha_1, \beta \right) = \ln 	\beta + \alpha_1 \bprp + \alpha_3 \feh + \alpha_4 \ln T_\mathrm{eff}
\label{eq:m5}
\end{equation}
In this version of the model we consider the dependency of $\tc$ with $\bprp$, $\feh$ and $\teff$. Again the introduction of metallicity somewhat makes the fitting process more difficult, slightly worsening the dispersion of the residuals and at the same time reinforcing the presence of some oscillatory trends. The situation is however significantly improved with respect to model $\mathcal{M}_3$ due to the introduction of temperature, upon which we suspect metallicity has some intrinsic dependency. Similar to the other models the fitting for samples $\sone$ and $\stwo$ alone are still significantly better then that obtained for sample $\sthree$.
 \begin{figure}[h]
   \centering
  \includegraphics[width=18cm]{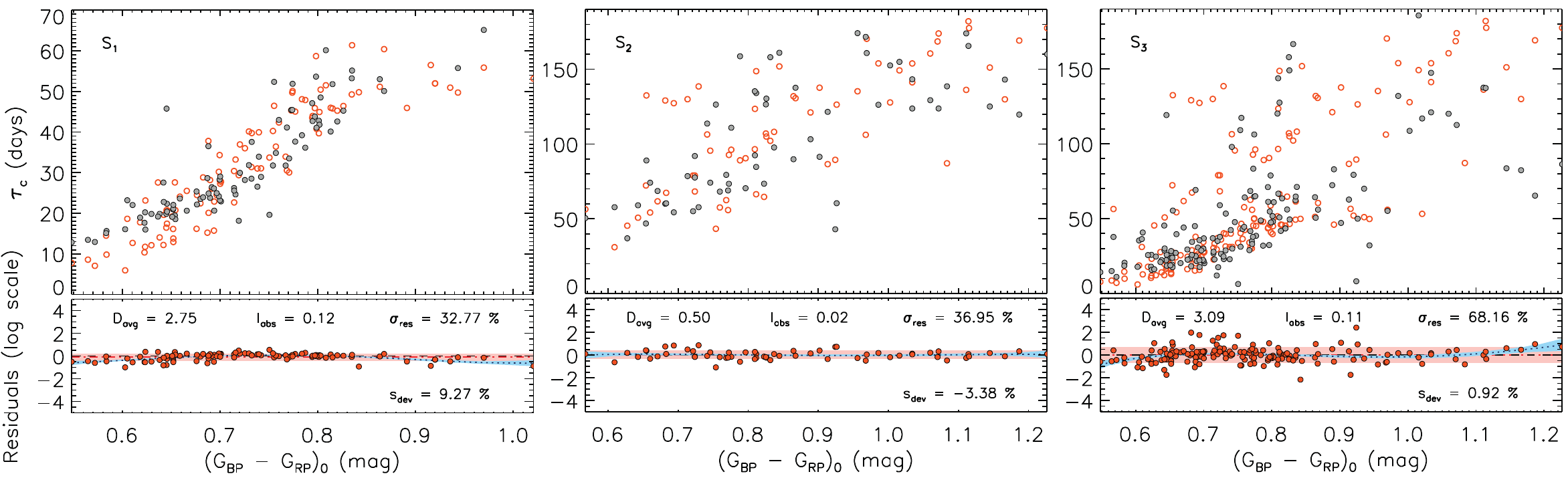}
  \includegraphics[width=18cm]{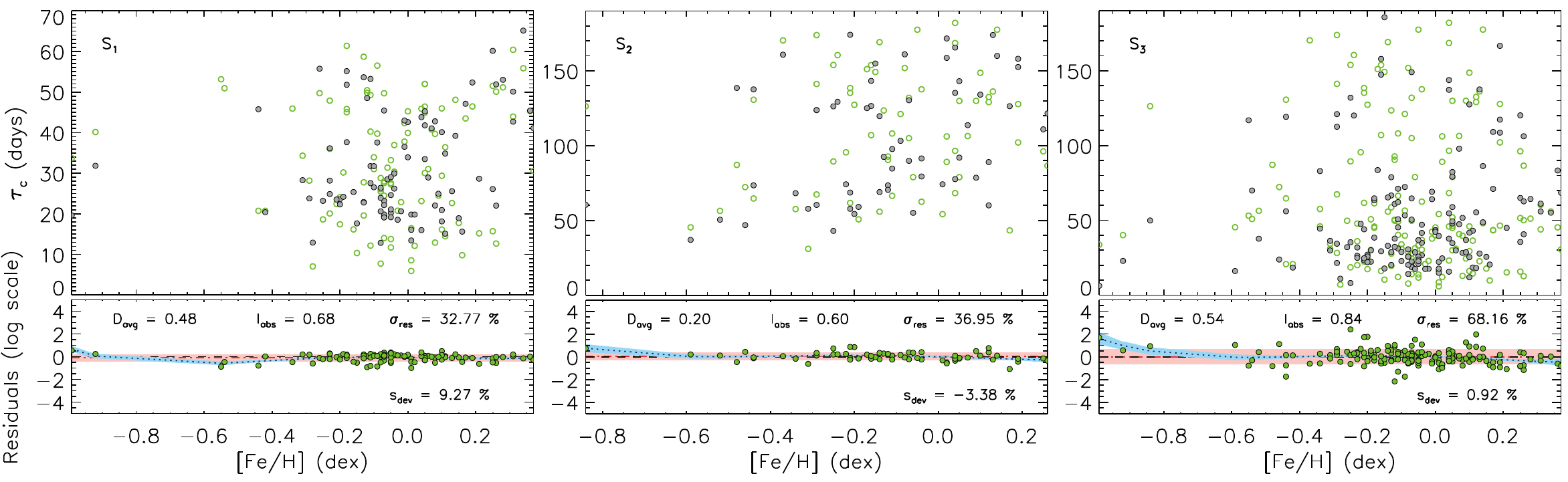}
    \includegraphics[width=18cm]{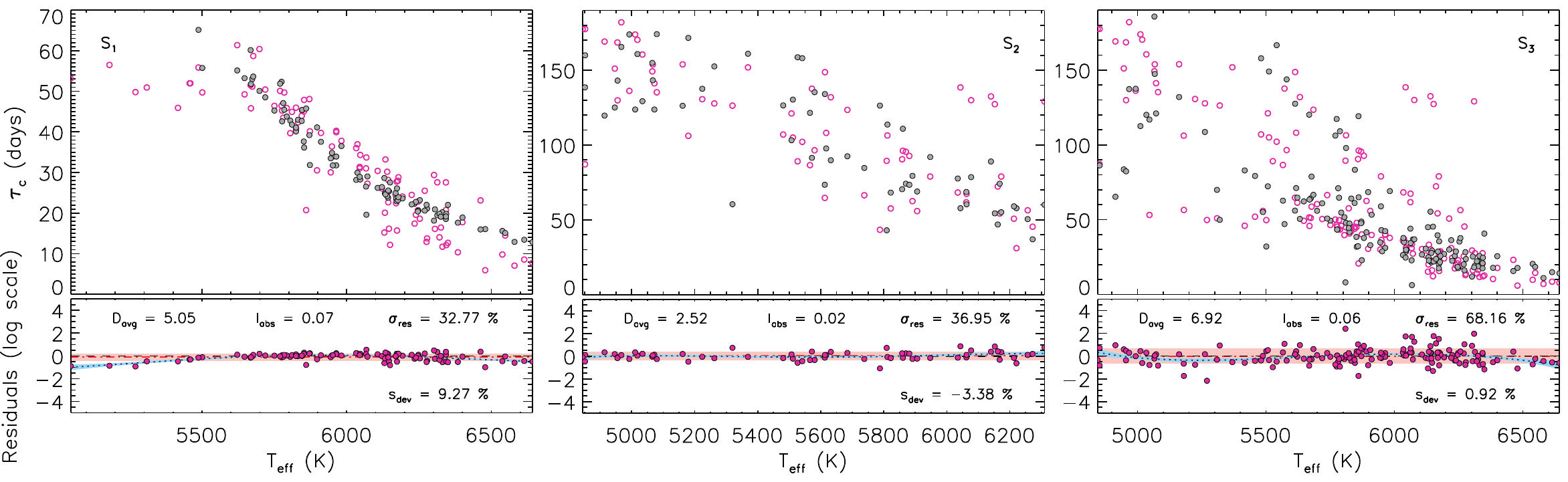}
    \caption{Similar description as for Fig.~\ref{fig:m1} but for model $\mathcal{M}_5$.}
    \label{fig:m5}
\end{figure}

\subsection{Model $\mathcal{M}_6$ ($\alpha_1 \neq 0$, $\alpha_2 \neq 0$, $\alpha_3 = 0$, $\alpha_4 \neq 0$)}
\begin{equation}
\ln \tc \left( \alpha_1, \beta \right) = \ln 	\beta + \alpha_1 \bprp + \alpha_2 \log g + \alpha_4 \ln T_\mathrm{eff}
\label{eq:m6}
\end{equation}
In this version of the model we consider the dependency of $\tc$ with $\bprp$, $\log g$ and $\teff$. Here surface gravity appears to be a relevant parameter for encompassing a good fraction of the dispersion seen in the samples that are considered. As a result $\sigma_\mathrm{res}$ is quite improved with respect to previous models. However, some clear issue is seen in the capabilty of the model to reproduce the temperature dependency in the sample $\sone$, while non-linear trends are still quite clear in the residuals.
 \begin{figure}[h]
   \centering
  \includegraphics[width=18cm]{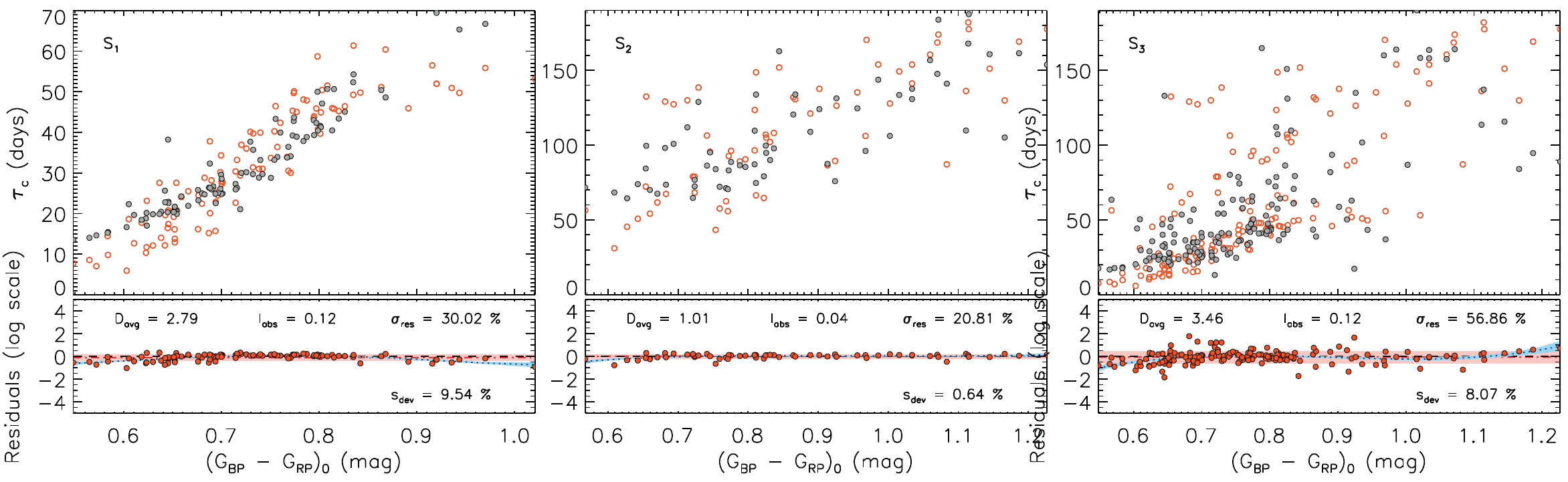}
  \includegraphics[width=18cm]{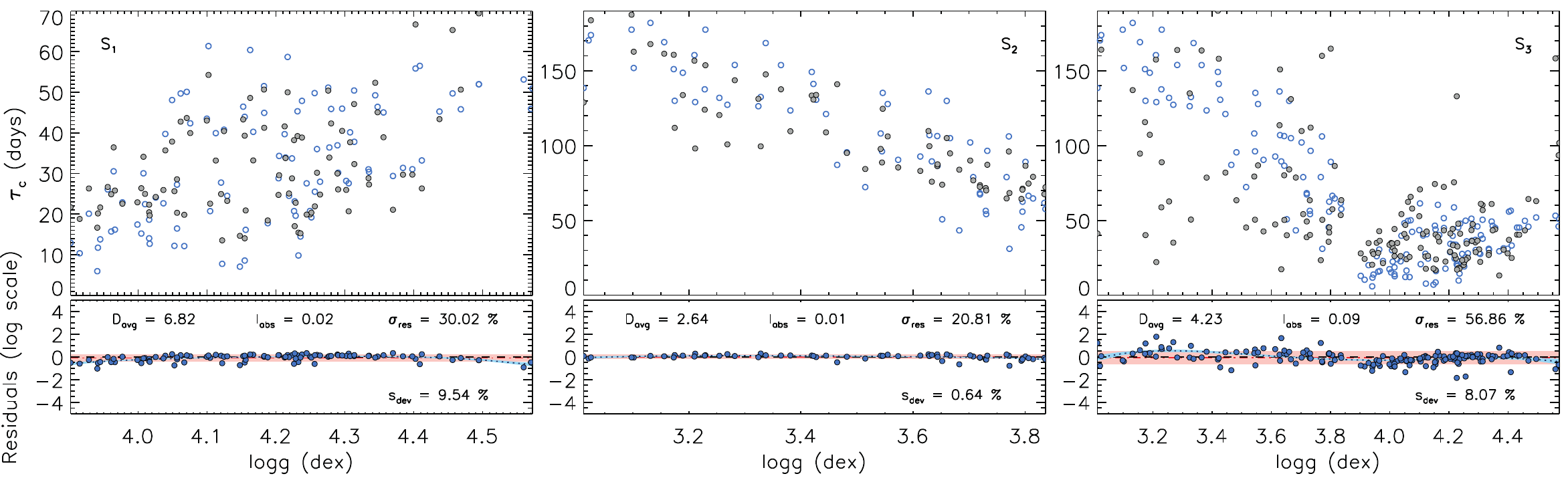}
   \includegraphics[width=18cm]{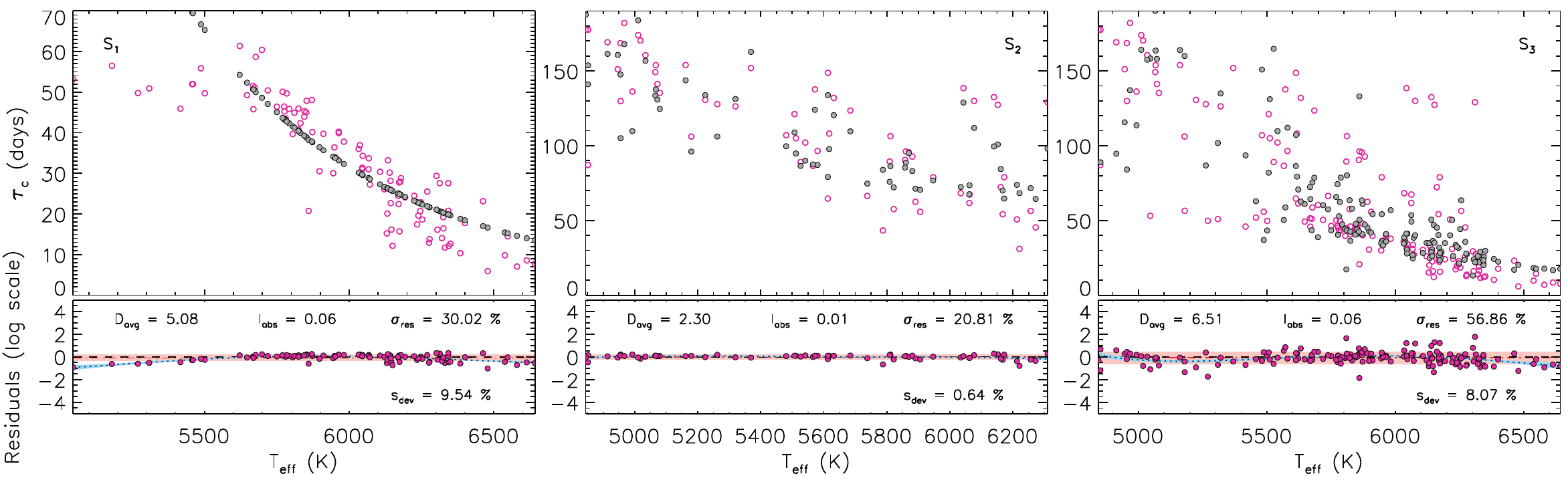}
    \caption{Similar description as for Fig.~\ref{fig:m1} but for model $\mathcal{M}_6$.}
    \label{fig:m6}
\end{figure}

\subsection{{Model $\mathcal{M}_7$ ($\alpha_1 \neq 0$, $\alpha_2 \neq 0$, $\alpha_3 \neq 0$, $\alpha_4 = 0$)}}
\begin{figure}[h]
   \centering
  \includegraphics[width=18cm]{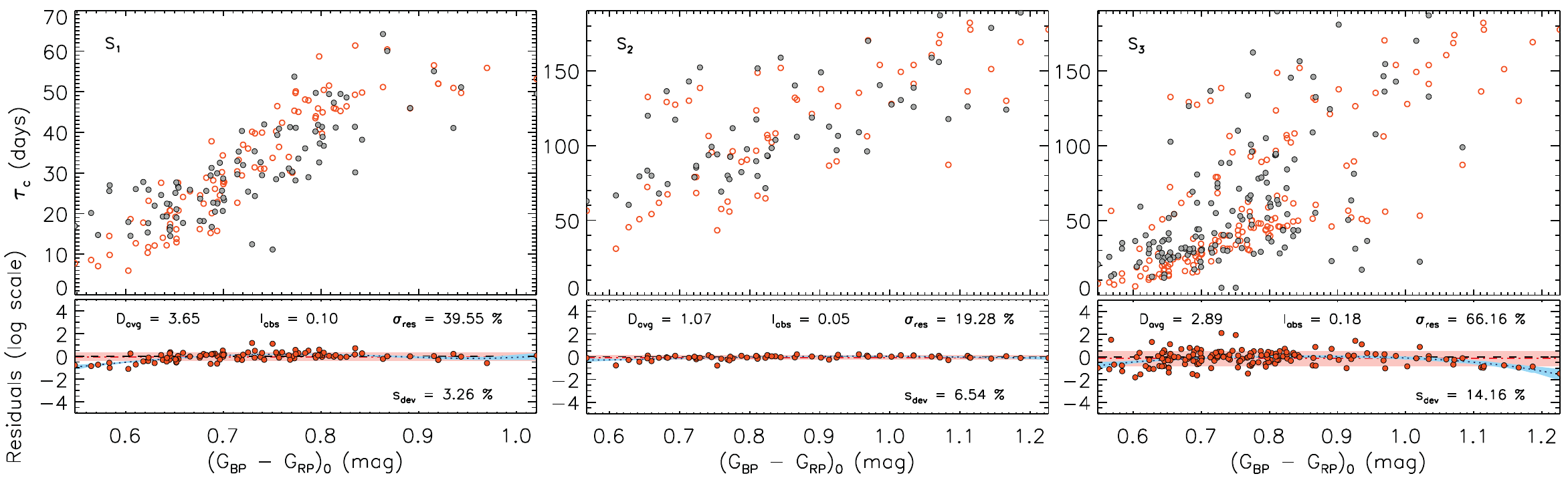}
  \includegraphics[width=18cm]{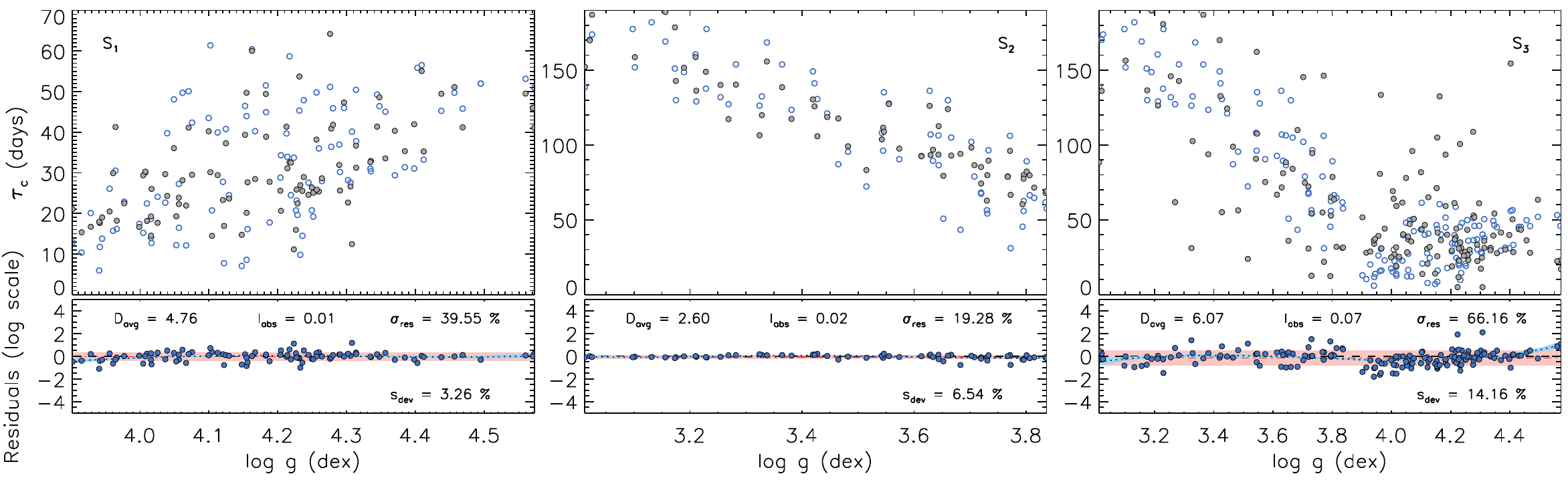}
  \includegraphics[width=18cm]{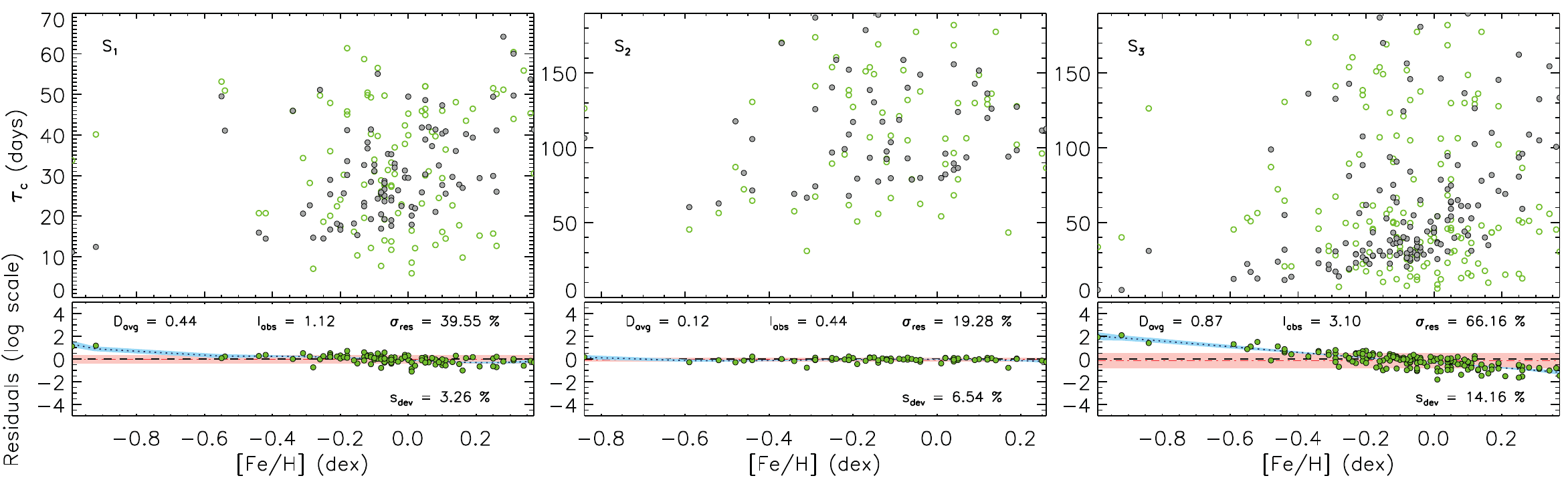}
    \caption{Similar description as for Fig.~\ref{fig:m1} but for model $\mathcal{M}_7$.}
    \label{fig:m7}
\end{figure}

\begin{equation}
\ln \tc \left( \alpha_1, \beta \right) = \ln 	\beta + \alpha_1 \bprp + \alpha_2 \log g + \alpha_3 \feh
\label{eq:m7}
\end{equation}
{In this version of the model we consider the dependency of $\tc$ with $\bprp$, $\log g$, $\feh$. The results for the model are shown in Fig.~\ref{fig:m7} for all the samples. Some pronounced trends are still visible in relation to the metallicity $\feh$, and especially for samples $\sone$ and $\sthree$, where the models appear to miss some relevant information. Here the trends with metallicity are worse than the case of model $\mathcal{M}_5$ because $\teff$ information is missing in this model, once again suggesting the presence of an intrinsic correlation between $\feh$ and $\teff$. Overall, the dispersion of the residuals is also worse than that of model $\mathcal{M}_6$, where the metallicity term is not included because it is replaced by a term with $\teff$.}

\clearpage
\subsection{Model $\mathcal{M}_8$ ($\alpha_1 \neq 0$, $\alpha_2 \neq 0$, $\alpha_3 \neq 0$, $\alpha_4 \neq 0$)}
\label{sec:m8}
\begin{figure}[h]
   \centering
  \includegraphics[width=18cm]{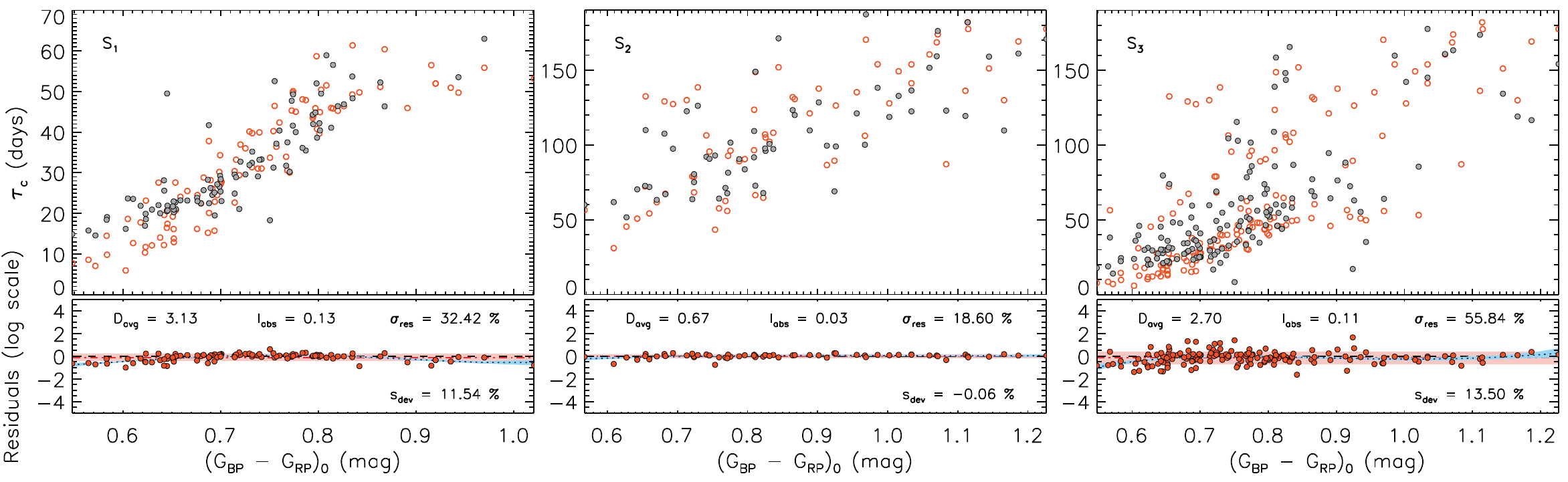}
  \includegraphics[width=18cm]{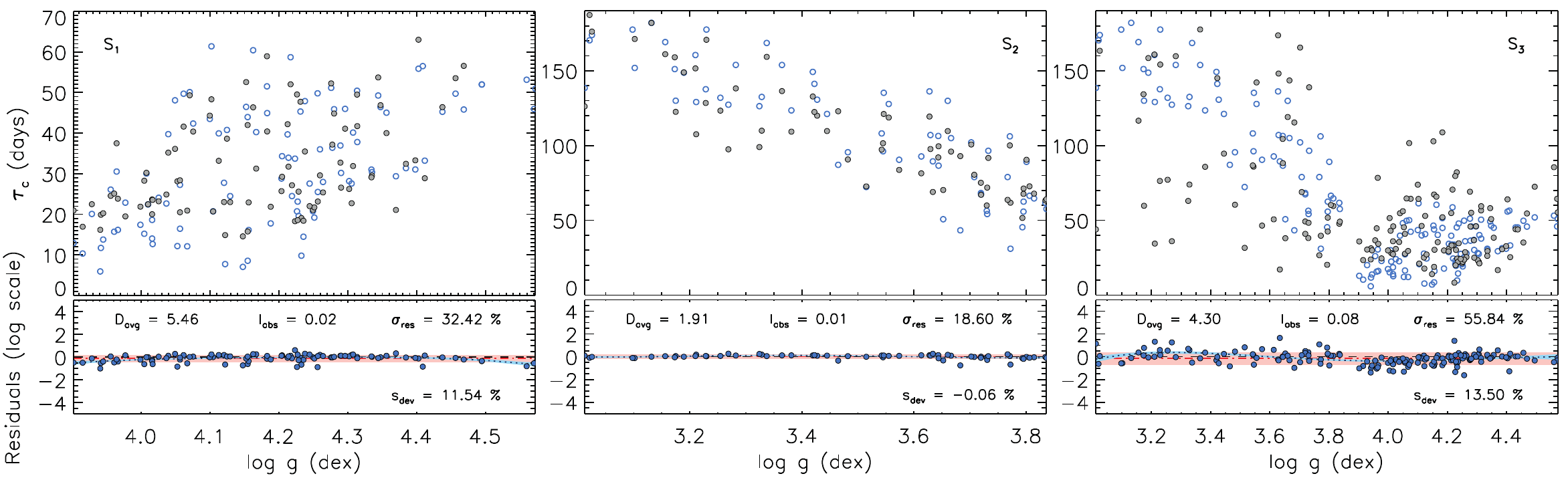}
  \includegraphics[width=18cm]{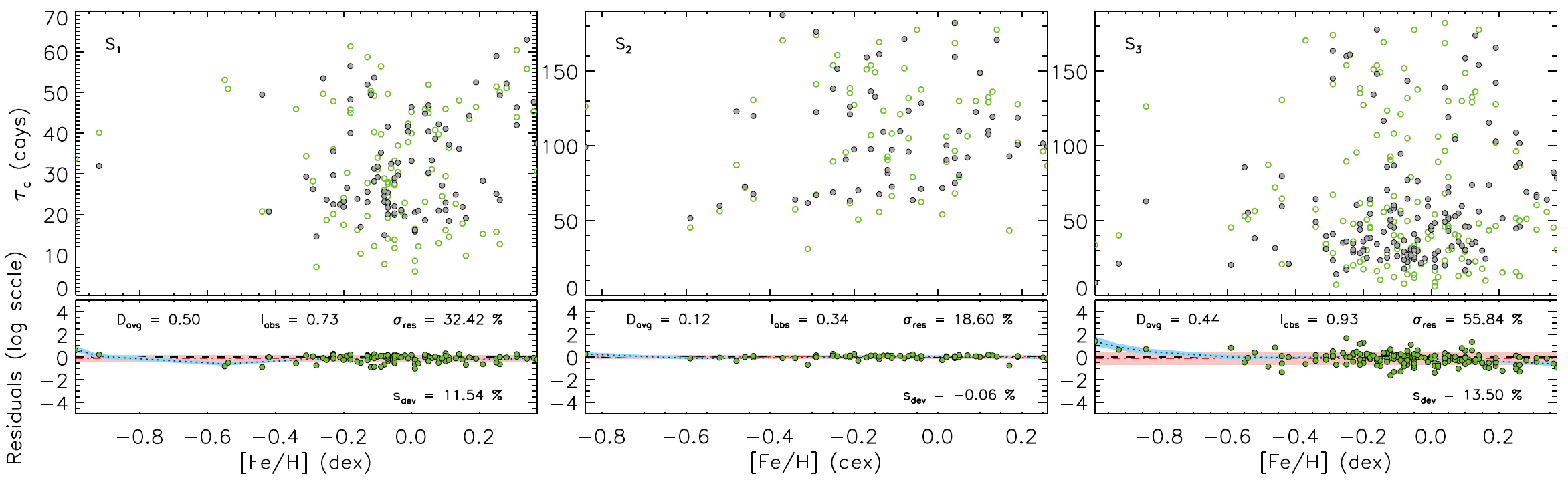}
  \includegraphics[width=18cm]{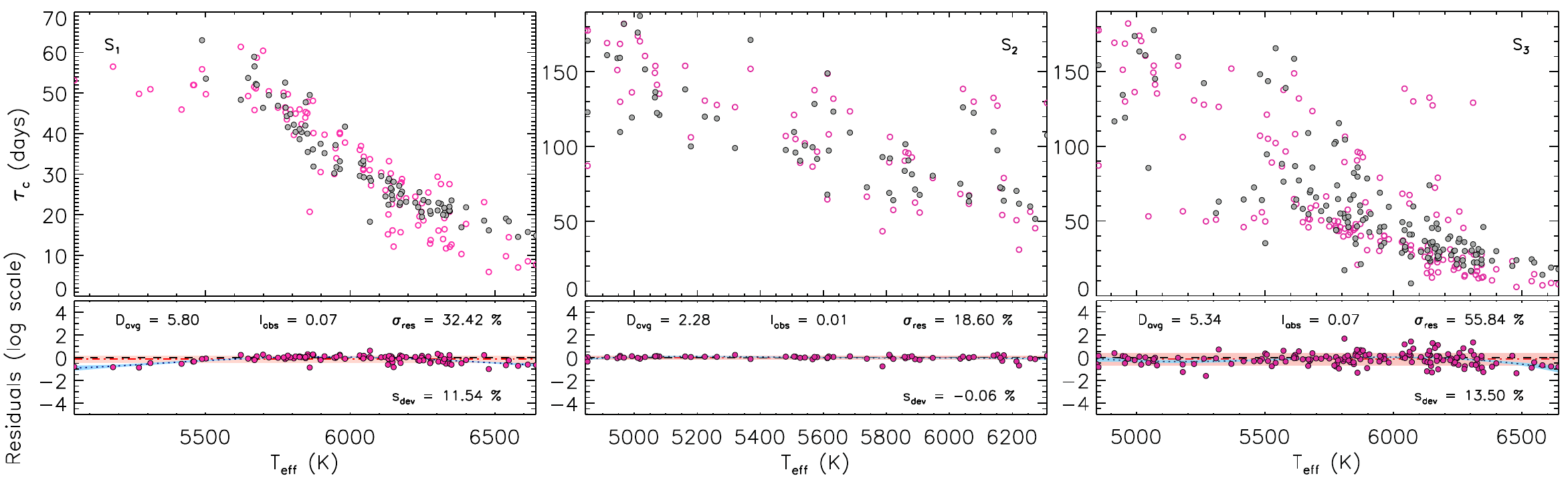}
    \caption{Similar description as for Fig.~\ref{fig:m1} but for model $\mathcal{M}_8$.}
    \label{fig:m8}
\end{figure}

\begin{equation}
\ln \tc \left( \alpha_1, \beta \right) = \ln 	\beta + \alpha_1 \bprp + \alpha_2 \log g + \alpha_3 \feh + \alpha_4 \ln T_\mathrm{eff}
\label{eq:m8}
\end{equation}
In this version of the model we consider the dependency of $\tc$ upon all the observables, namely $\bprp$, $\log g$, $\feh$, and $\teff$. The results for the model are shown in Fig.~\ref{fig:m8} for all the samples. The matching is significantly improved with respect to previous cases, and although some oscillatory trends still remain their magnitude is appreciably reduced and in line with the dispersion of the residuals, suggesting that this indeed represents a more suitable model to reproduce the observations. Despite the inclusion of metallicity alone in previous models had the net effect of worsening the results, in this case its contribution in conjuction with temperature and surface gravity appears to lead to an improvement with respect to model $\mathcal{M}_6$ where $\feh$ was not considered instead. Similar to all the previous cases, the combined sample $\sthree$ yields the worst result among the three inspected, an indication that the two samples $\sone$ and $\stwo$ might be difficult to be treated as a whole by using a common power-law model.

\clearpage
\begin{table*}
\small
\centering
 \caption{Parameter estimates for the favored power-law models according to the Bayesian model comparison as obtained by \textsc{Diamonds}, for each sample investigated. Median values and corresponding 68.3\,\% Bayesian credible limits are indicated for each free parameter.}
\begin{tabular}{lcrrrrr}
  \hline
  \\[-8pt]
Sample & Model & \multicolumn{1}{c}{$\alpha_1$} & \multicolumn{1}{c}{$\alpha_2$} & \multicolumn{1}{c}{$\alpha_3$}& \multicolumn{1}{c}{$\alpha_4$} & \multicolumn{1}{c}{$\ln \beta$}\\[1pt]
  \hline
  \\[-8pt]
$\sone$ & $\mathcal{M}_8$& $-3.10^{+      0.81}_{-0.60}$ & $0.55^{+0.12}_{-0.12}$ & $0.41^{+0.10}_{-0.10}$ & $-12.71^{+1.44}_{-1.26}$ & $114.00^{+10.83}_{-13.02}$ \\[1pt]
$\stwo$ & $\mathcal{M}_8$& $-0.93^{+0.31}_{-0.32}$ & $-0.89^{+0.06}_{-0.06}$ & $0.41^{+0.05}_{-0.06}$ & $-3.71^{+0.70}_{-0.58}$ & $40.59^{+5.35}_{-5.99}$ \\[1pt]
$\sthree$ & $\mathcal{M}_8$& $-7.82^{+1.01}_{-0.92}$ & $-0.37^{+0.06}_{-0.05}$ & $1.22^{+0.13}_{-      0.14}$ & $-21.86^{+1.99}_{-1.71}$ & $201.14^{+15.52}_{-18.23}$ \\[1pt]
   \hline
 \end{tabular}
\label{tab:coeff}
\end{table*}

\begin{table}
\small
\centering
 \caption{Correlation coefficients estimated from the posterior probability distribution obtained with \textsc{Diamonds} for the free parameters of the best model $\mathcal{M}_8$ of the three samples investigated.}
\begin{tabular}{lrrr}
  \hline
  \\[-8pt]
 & $\sone$ & $\stwo$ & $\sthree$ \\[1pt]
  \hline
  \\[-8pt]
\multicolumn{1}{l}{$\alpha_1$ vs. $\alpha_2$} & -0.50 & 0.30 & -0.60 \\[1pt]
\multicolumn{1}{l}{$\alpha_1$ vs. $\alpha_3$} & -0.22 & -0.47  & -0.41 \\[1pt]
\multicolumn{1}{l}{$\alpha_1$ vs. $\alpha_4$} & 0.92   & 0.95  & 0.98 \\[1pt]
\multicolumn{1}{l}{$\alpha_1$ vs. $\ln \beta$} & -0.92 & -0.96 & -0.98 \\[1pt]
\multicolumn{1}{l}{$\alpha_2$ vs. $\alpha_3$} & 0.26   & -0.17 & 0.07 \\[1pt]
\multicolumn{1}{l}{$\alpha_2$ vs. $\alpha_4$} & -0.28 & 0.11 & -0.64 \\[1pt]
\multicolumn{1}{l}{$\alpha_2$ vs. $\ln \beta$} & 0.25  & -0.16 & 0.63 \\[1pt]
\multicolumn{1}{l}{$\alpha_3$ vs. $\alpha_4$} & -0.16 & -0.51 & -0.44 \\[1pt]
\multicolumn{1}{l}{$\alpha_3$ vs. $\ln \beta$} & 0.15  & 0.51 & 0.44 \\[1pt]
\multicolumn{1}{l}{$\alpha_4$ vs. $\ln \beta$} & -1.00 & -1.00 & -1.00 \\[1pt]
   \hline
 \end{tabular}
\label{tab:corr}
\end{table}

\section{Discussion}
\label{sec:discussion}

\subsection{Statistical inference}
From a statistical point of view, independently of the sample adopted the inclusion of each observable is fully justified with a significant (many orders of magnitude) improvement in the corresponding Bayesian Evidence. Despite not all the observables considered are by definition independent quantities (e.g. the color index is directly, but not linearly, correlated with temperature), they constitute totally independent measurements coming from separate observations/instruments. By means of a multi-linear model that takes into account all the four observables used in this work, i.e. $\bprp$, $\log g$, $\feh$, $T_\mathrm{eff}$, the overall dispersion of the residuals is reasonably reduced, while systematics and trends therein appear to be contained within the observed scatter for the most part. Therefore, model $\mathcal{M}_8$ is, by far, the most favored one in this model comparison set and, at least from a statistical point of view, it is the one that ought to be preferred at predicting $\tc$ when $\bprp$, $\log g$, $\feh$, and $\teff$ are all available measurements. We note that $\teff$ has the largest statistical contribution when dealing with the sample $\sone$ of less evolved stars (see model $\mathcal{M}_2$), which is reflected in the extended sample $\sthree$, also containing the less evolved stars. {This can be explained by the narrower temperature range covered by giant stars with respect to the much wider one of the dwarfs.} For the sample $\stwo$ of more evolved stars instead, the major contribution arises from the surface gravity $\log g$ (see model $\mathcal{M}_4$), likely reflecting the large variation of $\log g$ that characterizes these targets.

In relation to metallicity, although its inclusion alone as {separate} from $\teff$ has the tendency to worsen the quality of the resulting fits, when it is instead incorporated with $\teff$ and $\log g$ it has the effect of improving the overall quality of the result, yielding a Bayesian evidence that is strongly improved. {From our evaluation of the correlation coefficients of the free parameters of model $\mathcal{M}_8$ listed in Table~\ref{tab:corr}, it is apparent that the free parameters referring to the dependency upon metallicity ($\alpha_3$) and temperature ($\alpha_4$) are not strongly correlated to one another irrespective of the sample considered. This reinforces the outcome of our model comparison process that the metallicity term is contributing with a significant and independent effect on the estimated convective turnover time. In particular, the positive values of $\alpha_3$ for $\sone$, $\stwo$, and $\sthree$ indicate not only that higher metallicities yield larger values of $\tau_c$ throughout different stages of stellar evolution, but also that this result is qualitatively in agreement with the fact that higher metallicities produce deeper convective zones because of the increased opacity, resulting in larger convective turnover times. It is also interesting to notice that the anti-correlation between $\alpha_3$ and $\alpha_4$, and similarly between $\alpha_3$ and $\alpha_1$, becomes more pronounced for samples $\stwo$ and $\sthree$, where giant stars are included. A likely explanation of this worsening when moving from dwarfs to giants can be given by the larger degree of degeneracy that is found in the fundamental properties of RGB stars with respect to dwarfs, because stars with different masses and metallicities all converge along the RGB, covering only a small range in effective temperature.}

{For what concerns the role of the evolutionary stage of the star,} a global effect is that the quality of the fit is lower when the two samples $\sone$ and $\stwo$ are combined, which can be observed for all models investigated. {Furthermore we notice a significant increase of the correlations among most of the free parameters of model $\mathcal{M}_8$ when moving from samples $\sone$, $\stwo$ to the full sample $\sthree$. This enhances the differences between dwarf and giant stars, and suggests that a common relationship is indeed less adequate at predicting $\tc$ than the two separate ones that are better tailored for each evolutionary stage (either MS or RGB). This result could potentially indicate} that the underlying physics adopted for obtaining the approximated definition of $\tc$ from \cite{Corsaro21} may be subject to changes as the star evolves from MS to RGB. Apparently, surface gravity, which is the parameter changing the most between the samples $\sone$ and $\stwo$ (see also Fig.~\ref{fig:correlations}) is not enough to properly reproduce the differences and reduce the residuals at a level comparable to that of the two samples taken individually. {The type of dependency with $\log g$ for samples $\sone$ and $\stwo$ is so different that, on top of what can already be seen in Fig.~\ref{fig:correlations} from the observables alone (i.e. the opposite sign of the Spearman's rank correlation coefficient for $\tc$ vs $\log g$), it is additionally reflected in the reversal of the correlation between $\alpha_2$ and $\alpha_3$, and between $\alpha_2$ and $\alpha_4$ when going from dwarfs to giants.} On one hand, if optimization is the priority, we suggest to consider separate power-law relations depending on whether the star is in or close to the MS stage of evolution (hence including early SG stars), or already beyond it (i.e. late SG and early RGB stars). In order to distinguish whether one or the other power-law could be adopted, one may apply a simple cut in $\log g$ given by the one seen in Fig.~\ref{fig:correlations} for $\tc$ vs. $\log g$ (taking the midpoint between the two samples), yielding $\log g \simeq 3.869$. One has to note, however, that this cut is strictly dependent on the specific catalogs adopted in this work. On the other hand, a common power-law relation is, from a physics point of view, a stronger outcome of this analysis, which could still be provided and possibly improved in the future by incorporating additional dependencies that can affect convection in stars (e.g. rotation). 

Another {aspect is related} to the striking good quality of the fits when the sample $\stwo$ alone is taken into account. As clearly seen from the residuals of most fits, here the models are capable of reproducing even more detailed sub-structures inside the cloud of observations. This not only is an indication that a power-law relation is indeed an adequate model for such observations, but also that the adopted measurements of $\log g$, $\feh$, and $\teff$ may appear to be rather robust against systematics and trends caused by any internal correlations among these quantities. Conversely, a similar discussion cannot be done for the sample $\sone$, where residuals are visibly larger for all models if compared to those of the sample $\stwo$. On top of a larger dispersion, the presence of unfitted structures in the residuals and the problems caused when introducing $\feh$ alone (visible especially for model $\mathcal{M}_3$), are an indication that some flaws could be present in the set of measurements adopted for this sample (e.g. about the correlation that could be present between $\feh$ and $\teff$ for MS stars). Finally, the higher level of dispersion that is seen in all the residuals of the sample $\sthree$ as compared to that of $\sone$ and $\stwo$, should not be surprising. This is because $\sthree$ carries all the potential issues already present in $\sone$ with the addition of an increased difficulty at predicting $\tc$ because of the adoption of a more numerous and diverse set of observations, implying a wider range of parameter values involved in the fits.

{The strong anti-correlation between $\alpha_4$ (and similarly $\alpha_1$) and the offset term $\ln \beta$ reported in Table~\ref{tab:corr} can be explained as the result of a fit balance between the offset and the temperature term. The temperature term is not only the strongest in absolute terms (given that the factors appearing in Eq.~(\ref{eq:multi-linear_2}) are not normalized), but also the one subject to the smallest relative variation (the overall temperature range spanned by our full sample is of nearly 2000\,K, resulting in about 28\,\% variation of this observable), hence behaving more similarly to a nearly constant term. Despite the mentioned striking anti-correlation, we have tested that a model $\mathcal{M}_8$ deprived of its offset term $\ln \beta$ is strongly disfavored from a statistical point of view because its Bayesian evidence is lower than that of our best model by many orders of magnitude for all the samples investigated.}

One last point concerns our comparison to the quadratic law identified by \cite{Corsaro21}, i.e. the model $\mathcal{M}_\mathrm{quadratic}$, as already anticipated in Sect.~\ref{sec:multilinear}. The results of our model comparison process show that in the case of samples $\sone$ and $\stwo$, $\mathcal{M}_\mathrm{quadratic}$ is significantly worse at predicting $\tc$ than all the eight multi-linear models considered in this work, while for the sample $\sthree$ $\mathcal{M}_\mathrm{quadratic}$ does a better job than model $\mathcal{M}_1$ because of the more complex structures arising when combining the $\sone$ and $\stwo$ together, but still remains significantly worse than all the remaining models considered. This result thus confirms that the multi-linear relations investigated here provide a solid extension and a substantial improvement to the previous work that was confined only to MS stars. 

\subsection{Connection to stellar dynamo}
It is important to emphasize that our prescription for $\tc$ should effectively capture the relevant time scales of turbulent eddies, even in stars with shallower convective zones. This capability is crucial for interpreting the Rossby number in the context of dynamo action.
Although the precise dynamo mechanism operating in solar-like stars remains uncertain, two primary mechanisms are considered within the framework of mean-field dynamo theory. If the meridional circulation is sufficiently strong, the dynamo cycle period is largely determined by the topology of the meridional flow at the base of the convection zone. Conversely, in a classical $\alpha\Omega$ dynamo mechanism, the characteristic cycle period is primarily governed by the efficiency of differential rotation, particularly the radial shear, which is prominent near the base of the convection zone. Attempts to model the solar dynamo using subsurface shear have produced topologies that are inconsistent with current observations \citep{2013SoPh..287..185B}. Additionally, an $\alpha$-effect uniformly distributed throughout the entire convection zone tends to yield dynamo solutions with incorrect parity \citep{2002A&A...390..673B}.

For these reasons, it is reasonable to propose that even very short dynamo cycles, such as that observed in the F8V star Iota Horologii \citep{Metcalfe_2010}, can be explained within the framework of a deep-seated $\alpha\Omega$ dynamo. Let us focus on this case and consider  an $\alpha$-effect acting close to the bottom of the convection zone and a (standard) magnetic turbulent diffusivity $\eta$ taken to be uniform in the upper part of the convection zone with a sudden drop in the transition region towards the radiation zone. Let us also assume an internal differential rotation of the solar like type.

\begin{figure}[ht]
   \centering
  \includegraphics[width=\textwidth]{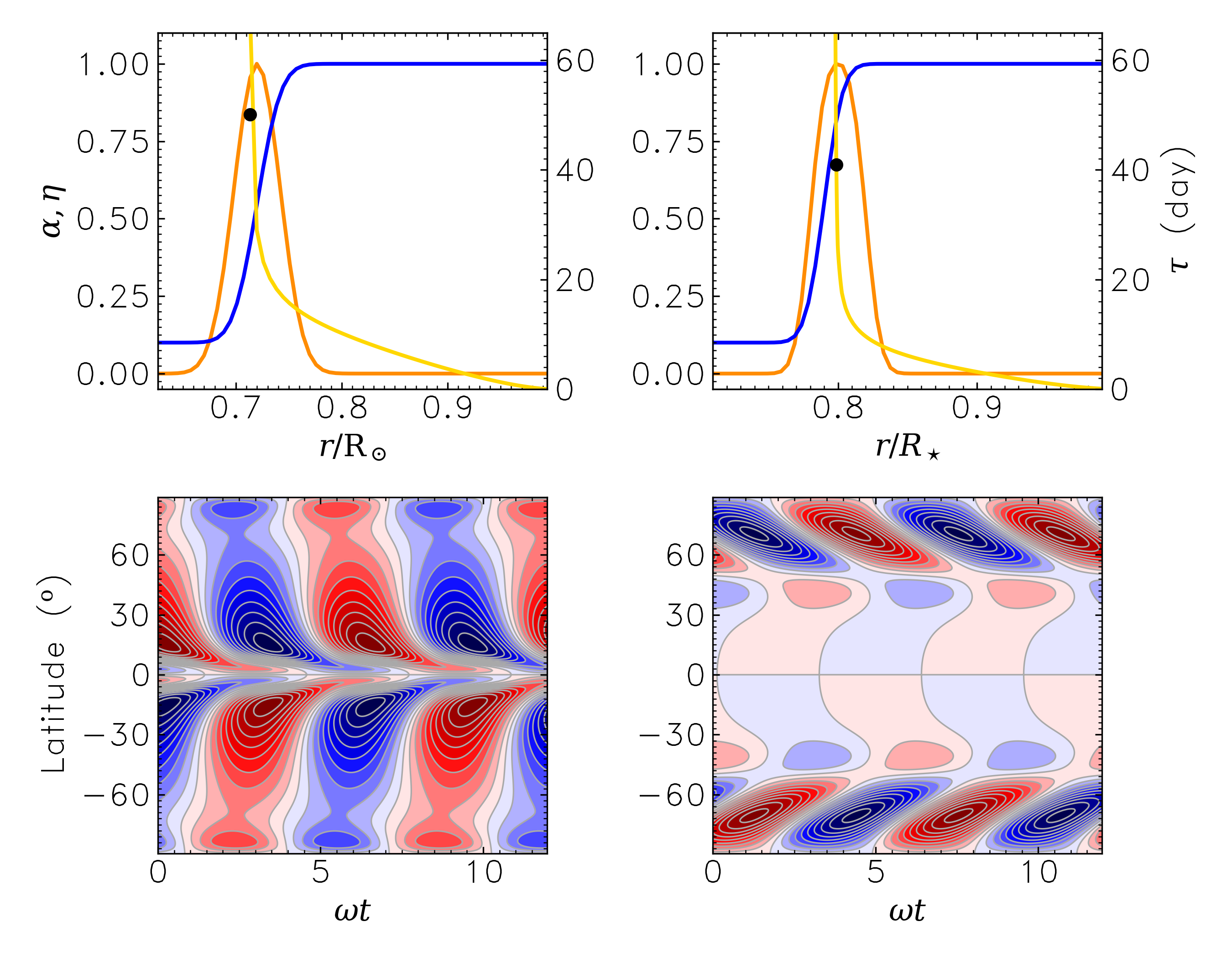}
    \caption{Dynamo model obtained for the Sun (\textit{left}) and iota Horologii (\textit{right}). In each case, the top panel shows the magnetic diffusivity profile $\eta$ (blue), the $\alpha$-effect profile (orange), and the convective velocity (yellow). The black dots highlight at which the $\tc$ profile intersects the convective turnover time computed with the prescription from \citet{Corsaro21}. 
    The bottom panels shows the evolution of the mean toroidal field obtained for each model, at $r=0.78$~$\rm R_\odot$ and $r=0.8$~$R_\star$, respectively. The magnetic field in the regions in red has an opposite sign with respect to the regions in blue.
    }
    \label{fig:tauprofile}
\end{figure}

The results, based on the mean field dynamo code CTDYN \citep{2008A&A...483..949J} are shown in Fig.~(\ref{fig:tauprofile}).  The panels on the top represent the radial profile of the $\alpha$ effect and of the magnetic diffusivity profile $\eta$ that were used in the CTDYN computation for the Sun (left) and for a F8V star.
For actual calculations we used the Catania version of the GARSTEC code to extract the convective velocities and the convective turnover time $\tc$ as a function of the radial distance. The $\tc$ profile is computed according to   $\tc = \alpha_{\rm MLT} H_p / v_c $
where $\alpha_{\rm MLT} = 1.65$ is the mixing-length coefficient, $H_p$ the pressure scale height, and $v_c$ the convective velocity.
In the bottom panel, we show the temporal evolution of the mean toroidal magnetic field at $r=0.78$~$\rm R_\odot$ and $r=0.8$~$R_\star$, respectively. 
The properties of the dynamo mechanisms are controlled by the dimensionless numbers $C_\Omega = \Omega_\star R_\star^2 / \eta$, $R_m$, and $C_\alpha$, which quantify the importance of the rotation, the meridional circulation, and the $\alpha$ effect, respectively \citep{2008A&A...483..949J}. 
$C_\Omega$ and $C_u$ are provided as model input while the corresponding critical $C_\alpha$ value is searched for by CTDYN when solving the eigenmode problem.
We set $C_\Omega = 3\times10^4$ and $R_m=400$ in the case of the Sun while we  take $C_\Omega = 1\times10^3$ and $R_m=0$ for the F8V model (that is no meridional circulation).
We obtain from the CTDYN computation $C_\alpha = 9.1$ for the Sun and $C_\alpha = 27.9$ for the FV8 model.
The angular frequency $\omega$ of the Hale-like cycle (the polarity of the field returns to its original state) that we obtain correspond to a period of 20.3~years in the solar model and of 3.3~years for the F8V model, confirming that having a deep-seated $\alpha$ effect allows us to reproduce both long and short periods cycle. We also note that, while the solar model exhibit the well-known butterfly-like behaviour, the magnetic field in the case of the F8V star is concentrated at high latitude, close to the poles. 

\section{Conclusions}
\label{sec:conclusions}
The results of our analysis indicate that the adoption of a multi-linear relationship for predicting $\tc$ that take into account the observables $\bprp$, $\teff$, $\log g$, $\feh$, yields significantly improved predictions with respect to the quadratic law presented by \cite{Corsaro21}, who only adopted a dependence upon color index and a more limited sample comprising essentially MS stars. In this work, we have extended the range of applicability of such asteroseismic calibration to stars having evolutionary stages beyond the MS and up to the RGB phase. In particular, we calibrated new relationships that can cover all these stages of evolution either with one common law, or by separating them between MS and more evolved stars with the advantage of reducing the dispersion in the residuals, thus leading to a better overall precision. Care remains to be given to the specific measurements adopted for metallicity when predicting $\tc$, which we found to be at least partially correlated to $\teff$ in the sample of MS stars. 

It is interesting to note that our favored model does a rather good job at reproducing the fine structure that can be seen in the values of $\tc$, including the possible flattening happening toward redder colors in the case of MS stars. This implies that simple linear dependencies upon the observables adopted already suffice at encompassing with reasonable accuracy the main behavior observed in $\tc$ as a function of stellar properties, without the need to invoke for a more complex quadratic dependency. Also, despite $\bprp$ is clearly sensitive to $\teff$, they are not the same physical quantity, and our analysis suggests that both observables should be taken into account for improving the prediction on $\tc$.

\clearpage

\begin{table*}
\caption{Statistical indices assessing the quality of the fits for all the models considered in this work. For a definition of the different parameters, see Sect.~\ref{sec:results}.}             
\centering             
\tiny
\begin{tabular}{l c c c c c c c c c c c c c}
\hline
\\[-4pt]
\multirow{2}[2]*{Model} & \multirow{2}[2]*{Sample} & \multirow{2}[2]*{$s_\mathrm{dev}$ (\%)} & \multirow{2}[2]*{$\sigma_\mathrm{res}$ (\%)} & & \multicolumn{4}{c}{$D_\mathrm{avg}$} & & \multicolumn{4}{c}{$I_\mathrm{abs}$} \\[10pt]
\cmidrule{6-9}
\cmidrule{11-14}
\\[-8pt]
\multicolumn{5}{c}{}  & $\bprp$ & $\log g$ & $\feh$ & $\teff$ & & $\bprp$ & $\log g$ & $\feh$ & $\teff$\\[10pt]
\cmidrule{1-14}
\\[-8pt]
\multirow{3}[3]*{$\mathcal{M}_1$} & $\sone$ &19.22 & 33.92 & & 3.555 & - & - & - & & 0.129 & - & - & - \\ [1pt]
\cmidrule{2-4}
\cmidrule{6-9}
\cmidrule{11-14}
\\[-8pt]
& $\stwo$ & 10.82 & 29.91 & & 1.014 & - & - & - & & 0.068 & - & - & - \\ [1pt]
\cmidrule{2-4}
\cmidrule{6-9}
\cmidrule{11-14}
\\[-8pt]
& $\sthree$ & 10.64 & 54.15 & & 3.095 & - & - & - & & 0.098 & - & - & - \\ [1pt]
\cmidrule{1-14}
\\[-8pt]
\multirow{3}[3]*{$\mathcal{M}_2$} & $\sone$ & 12.08 & 33.23 & & 2.768 & - & - & 5.048 & & 0.133 & - & - & 0.074 \\ [1pt]
\cmidrule{2-4}
\cmidrule{6-9}
\cmidrule{11-14}
\\[-8pt]
& $\stwo$ & -2.40 & 36.97 & & 1.884 & - & - & 1.534 & & 0.056 & - & - & 0.012 \\ [1pt]
\cmidrule{2-4}
\cmidrule{6-9}
\cmidrule{11-14}
\\[-8pt]
& $\sthree$ & 10.43 & 64.63 & & 3.620 & - & - & 7.334 & & 0.143 & - & - & 0.068 \\ [1pt]
\cmidrule{1-14}
\\[-8pt]
\multirow{3}[3]*{$\mathcal{M}_3$} & $\sone$ & 7.23 & 41.05 & & 3.103 & - & 0.453 & - & & 0.089 & - & 1.420 & - \\ [1pt]
\cmidrule{2-4}
\cmidrule{6-9}
\cmidrule{11-14}
\\[-8pt]
& $\stwo$ & 9.68 & 41.20 & & 2.082 & - & 0.406 & - & & 0.083 & - & 1.506 & - \\ [1pt]
\cmidrule{2-4}
\cmidrule{6-9}
\cmidrule{11-14}
\\[-8pt]
& $\sthree$ & 33.36 & 153.97 & & 7.803 & - & 2.368 & - & & 0.304 & - & 9.245 & - \\ [1pt]
\cmidrule{1-14}
\\[-8pt]
\multirow{3}[3]*{$\mathcal{M}_4$} & $\sone$ & 20.56 & 36.67 & & 3.318 & 9.170 & - & - & & 0.123 & 0.034 & - & - \\ [1pt]
\cmidrule{2-4}
\cmidrule{6-9}
\cmidrule{11-14}
\\[-8pt]
& $\stwo$ & 6.24 & 21.23 & & 1.150 & 2.332 & - & - & & 0.047 & 0.014 & - & - \\ [1pt]
\cmidrule{2-4}
\cmidrule{6-9}
\cmidrule{11-14}
\\[-8pt]
& $\sthree$ & 22.73 & 48.65 & & 4.756 & 6.750 & - & - & & 0.157 & 0.097 & - & - \\ [1pt]
\cmidrule{1-14}
\\[-8pt]
\multirow{3}[3]*{$\mathcal{M}_5$} & $\sone$ & 9.28 & 32.77 & & 2.755 & - & 0.480 & 5.052 & & 0.121 & - & 0.678 & 0.067 \\ [1pt]
\cmidrule{2-4}
\cmidrule{6-9}
\cmidrule{11-14}
\\[-8pt]
& $\stwo$ & -3.38 & 36.95 & & 0.502 & - & 0.204 & 2.516 & & 0.023 & - & 0.603 & 0.017 \\ [1pt]
\cmidrule{2-4}
\cmidrule{6-9}
\cmidrule{11-14}
\\[-8pt]
& $\sthree$ & 0.92 & 68.16 & & 3.090 & - & 0.540 & 6.923 & & 0.110 & - & 0.835 & 0.057 \\ [1pt]
\cmidrule{1-14}
\\[-8pt]
\multirow{3}[3]*{$\mathcal{M}_6$} & $\sone$ & 9.54 & 30.02 & & 2.790 & 6.824 & - & 5.081 & & 0.123 & 0.024 & - & 0.064 \\ [1pt]
\cmidrule{2-4}
\cmidrule{6-9}
\cmidrule{11-14}
\\[-8pt]
& $\stwo$ & 0.64 & 20.81 & & 1.006 & 2.641 & - & 2.302 & & 0.040 & 0.015 & - & 0.012 \\ [1pt]
\cmidrule{2-4}
\cmidrule{6-9}
\cmidrule{11-14}
\\[-8pt]
& $\sthree$ & 8.07 & 56.86 & & 3.455 & 4.234 & - & 6.511 & & 0.121 & 0.088 & - & 0.061 \\ [1pt]
\cmidrule{1-14}
\\[-8pt]
\multirow{3}[3]*{$\mathcal{M}_7$} & $\sone$ & 3.258 & 39.55 & & 3.648 & 4.757 & 0.438 & - & & 0.097 & 0.011 & 1.118 & - \\ [1pt]
\cmidrule{2-4}
\cmidrule{6-9}
\cmidrule{11-14}
\\[-8pt]
& $\stwo$ & 6.54 & 19.28 & & 1.066 & 2.603 & 0.119 & - & & 0.049 & 0.017 & 0.443 & - \\ [1pt]
\cmidrule{2-4}
\cmidrule{6-9}
\cmidrule{11-14}
\\[-8pt]
& $\sthree$ & 14.16 & 66.16 & & 2.892 & 6.073 & 0.866 & - & & 0.176 & 0.074 & 3.096 & - \\ [1pt]
\cmidrule{1-14}
\\[-8pt]
\multirow{3}[3]*{$\mathcal{M}_8$} & $\sone$ & 11.54 & 32.42 & & 3.131 & 5.464 & 0.499 & 5.796 & & 0.128 & 0.023 & 0.730 & 0.072 \\ [1pt]
\cmidrule{2-4}
\cmidrule{6-9}
\cmidrule{11-14}
\\[-8pt]
& $\stwo$ &-0.06 & 18.60 & & 0.677 & 1.906 & 0.123 & 2.281 & & 0.029 & 0.014 & 0.336 & 0.012 \\ [1pt]
\cmidrule{2-4}
\cmidrule{6-9}
\cmidrule{11-14}
\\[-8pt]
& $\sthree$ & 13.50 & 55.84 & & 2.700 & 4.296 & 0.439 & 5.342 & & 0.108 & 0.076 & 0.935 & 0.066 \\ [1pt]
\cmidrule{1-14}
\\[-8pt]
\end{tabular}
\label{tab:stats}
\end{table*}

\clearpage
        \begin{table}[ht]
    \footnotesize
        \begin{center}
         \caption{Natural logarithms of the Bayes' factor $\ln\mathcal{B}_{ij}$ for each pair of models ($\mathcal{M}_i$, $\mathcal{M}_j$) analyzed for the case of the sample $\mathcal{S}_1$ (MS and early SGs). The favored model is highlighted in gray shading.}
            \begin{tabular}{l *{8}{r}}
              \hline
			  \multicolumn{1}{c}{-}  & \multicolumn{1}{c}{$\mathcal{M}_1$} & \multicolumn{1}{c}{$\mathcal{M}_2$} & \multicolumn{1}{c}{$\mathcal{M}_3$} & \multicolumn{1}{c}{$\mathcal{M}_4$} & \multicolumn{1}{c}{$\mathcal{M}_5$} & \multicolumn{1}{c}{$\mathcal{M}_6$} & \multicolumn{1}{c}{$\mathcal{M}_7$} & \multicolumn{1}{c}{$\mathcal{M}_8$} \\
              \hline
              $\mathcal{M}_1$ & \multicolumn{1}{c}{--} & -280.8 & -144.4 & -97.9 & -282.9 & -263.4 & -183.4 & -289.7\\

              $\mathcal{M}_2$ & 280.8 & \multicolumn{1}{c}{--} & 136.4 & 182.9 & -2.1 & 17.5 & 97.5 & -8.8\\

              $\mathcal{M}_3$ & 144.4 & -136.4 & \multicolumn{1}{c}{--} & 46.5 & -138.5 & -119.0 & -39.0 & -145.3\\

              $\mathcal{M}_4$ & 97.9 & -182.9 & -46.5 & \multicolumn{1}{c}{--} & -185.0 & -165.5 & -85.5 & -191.8\\
              
              $\mathcal{M}_5$ & 282.9 & 2.1 & 139.0 & 185.0 & \multicolumn{1}{c}{--} & 19.6 & 99.6 & -6.7\\
              
              $\mathcal{M}_6$ & 263.4 & -17.5 & 119.0 & 165.5 & -19.6 & \multicolumn{1}{c}{--} & 80.0 & -26.3\\

              $\mathcal{M}_7$ & 183.4 & -97.5 & 39.0 & 85.5 &-99.6 & -80.0 & \multicolumn{1}{c}{--} & -106.3\\

   \rowcolor[gray]{0.9} $\mathcal{M}_8$ &  289.7 & 8.8 & 145.3 & 191.8 & 6.7 & 26.3 & 106.3 & \multicolumn{1}{c}{--} \\
            \end{tabular}
        \label{tab:s1}
        \end{center}
    \end{table}

    \begin{table}[ht]
    \footnotesize
        \begin{center}
         \caption{Similar description as for Table~\ref{tab:s1} but for sample $\stwo$.}
            \begin{tabular}{l *{8}{r}}
              \hline
			  \multicolumn{1}{c}{-}  & \multicolumn{1}{c}{$\mathcal{M}_1$} & \multicolumn{1}{c}{$\mathcal{M}_2$} & \multicolumn{1}{c}{$\mathcal{M}_3$} & \multicolumn{1}{c}{$\mathcal{M}_4$} & \multicolumn{1}{c}{$\mathcal{M}_5$} & \multicolumn{1}{c}{$\mathcal{M}_6$} & \multicolumn{1}{c}{$\mathcal{M}_7$} & \multicolumn{1}{c}{$\mathcal{M}_8$} \\
              \hline
              $\mathcal{M}_1$ & \multicolumn{1}{c}{--} & -207.0 & -160.4 & -273.4 & -253.6 & -296.6 & -300.3 & -324.8\\
              
              $\mathcal{M}_2$ & 207.0 & \multicolumn{1}{c}{--} & 46.6 & -66.4 & -46.6 & -89.6 & -93.3 & -117.8\\
              
              $\mathcal{M}_3$ & 160.4 & -46.6 & \multicolumn{1}{c}{--} & -113.0 & -93.2 & -136.2 & -139.9 & -164.4\\
              
              $\mathcal{M}_4$ & 273.4 & 66.4 & 113.0 & \multicolumn{1}{c}{--} & 19.8 & -23.3 & -26.9 & -51.4\\
              
              $\mathcal{M}_5$ & 253.6 & 46.6 & 93.2 & -19.8 & \multicolumn{1}{c}{--} & -43.1 & -46.7 & -71.3\\
              
              $\mathcal{M}_6$ & 296.6 & 89.6 & 136.2 & 23.3 & 43.1 & \multicolumn{1}{c}{--} & -3.6 & -28.2\\
              
              $\mathcal{M}_7$ & 300.3 & 93.3 & 139.9 & 26.9 & 46.7 & 3.6 & \multicolumn{1}{c}{--} & -24.6\\
              
\rowcolor[gray]{0.9} $\mathcal{M}_8$ & 324.8 & 117.8 & 164.4 & 51.4 & 71.3 & 28.2 & 24.6 & \multicolumn{1}{c}{--} \\
            \end{tabular}
            \label{tab:s2}
        \end{center}
    \end{table}

        \begin{table}[ht]
    \footnotesize
        \begin{center}
         \caption{Similar description as for Table~\ref{tab:s1} but for sample $\sthree$.}
            \begin{tabular}{l *{8}{r}}
              \hline
			  \multicolumn{1}{c}{-}  & \multicolumn{1}{c}{$\mathcal{M}_1$} & \multicolumn{1}{c}{$\mathcal{M}_2$} & \multicolumn{1}{c}{$\mathcal{M}_3$} & \multicolumn{1}{c}{$\mathcal{M}_4$} & \multicolumn{1}{c}{$\mathcal{M}_5$} & \multicolumn{1}{c}{$\mathcal{M}_6$} & \multicolumn{1}{c}{$\mathcal{M}_7$} & \multicolumn{1}{c}{$\mathcal{M}_8$} \\
              \hline
                            $\mathcal{M}_1$ & \multicolumn{1}{c}{--} & -4396.0 & -3928.0 & -3869.6 & -4427.8 & -4396.6 & -4176.8 & -4441.0\\
              
              $\mathcal{M}_2$ & 4396.0 & \multicolumn{1}{c}{--} & 468.1 & 526.4 & -31.8 & -0.5 & 219.2 & -44.9\\
              
              $\mathcal{M}_3$ & 3928.0 & -468.1 & \multicolumn{1}{c}{--} & 58.3 & -499.9 & -468.6 & -248.9 & -513.0\\
              
              $\mathcal{M}_4$ & 3869.6 & -526.4 & \-58.3 & \multicolumn{1}{c}{--} & -558.2 & -527.0 & -307.2 & -571.3\\
              
              $\mathcal{M}_5$ & 4427.8 & 31.8 & 499.9 & 558.2 & \multicolumn{1}{c}{--} & 31.3 & 251.0 & -13.1\\
              
              $\mathcal{M}_6$ & 4396.6 & 0.5 & 468.6 & 527.0 & 31.3 & \multicolumn{1}{c}{--} & 219.8 & -44.4\\
              
              $\mathcal{M}_7$ & 4176.8 & -219.2 & 248.9 & 307.2 & -251.0 & -219.8 & \multicolumn{1}{c}{--} & -264.1\\
              
\rowcolor[gray]{0.9}  $\mathcal{M}_8$ & 4441.0 & 44.9 & 513.0 & 571.3 & 13.1 & 44.4 & 264.1 & \multicolumn{1}{c}{--} \\
            \end{tabular}
         \label{tab:s3}
        \end{center}
    \end{table}

\begin{acknowledgements}
We are thankful to Tanda Li and Tim Bedding for their insights. We acknowledge support from the research grant “Unveiling the magnetic side of the Stars” (PI A. Bonanno) funded under the INAF national call for Fundamental Research 2023. A.B. and E.C. are funded by the European Union – NextGenerationEU RRF M4C2 1.1  n: 2022HY2NSX. "CHRONOS: adjusting the clock(s) to unveil the CHRONO-chemo-dynamical Structure of the Galaxy” (PI: S. Cassisi). T.S.M.\ is supported by NASA grant 80NSSC25K7563. A.B. acknowledges support from the DFG grant AR 355/13-1.
\end{acknowledgements}


\end{document}